\documentclass[onecolumn,amsmath,amssymb,superscriptaddress]{revtex4-1}

\usepackage{verbatim} 

\usepackage{graphicx}

\usepackage{amstext,amsmath,amssymb,amsfonts}
\usepackage{bm}

\usepackage{mathtools}
\DeclareMathAlphabet{\mathpzc}{OT1}{pzc}{m}{it}
\DeclareFontFamily{OT1}{pzc}{}
\DeclareFontShape{OT1}{pzc}{m}{it}{<-> s * [1.100] pzcmi7t}{}
\DeclareMathAlphabet{\mathpzc}{OT1}{pzc}{m}{it}

\usepackage{hyperref}

\usepackage{color}
\definecolor{lightblue}{rgb}{0.2,0.2,0.7}
\definecolor{darkblue}{rgb}{0,0.25,0.5}
\definecolor{redbrown}{rgb}{0.875,0.25,0.125}
\definecolor{darkgreen}{rgb}{0,0.5,0}

\newcommand{\bra}[1]{\ensuremath{\langle #1 \vert}}
\newcommand{\ket}[1]{\ensuremath{\vert #1  \rangle}}
\newcommand{\braket}[2]{\ensuremath{\langle  #1, #2  \rangle}}

\renewcommand{\b}[1]{\ensuremath{\mathbf{#1}}}

\newcommand{\tr}{\ensuremath{\text{tr}}}

\renewcommand{\d}{\ensuremath{\text{d}}}

\DeclareMathOperator{\arctanh}{arctanh}
\DeclareMathOperator{\sgn}{sgn}
\newcommand{\argmax}[1]{\underset{#1}{\text{argmax}}}

\newcommand{\isEquivTo}[1]{\underset{#1}{\sim}}

\renewcommand{\i}{\ensuremath{\text{i}}}

\newcommand{\calh}{\ensuremath{\mathpzc{h}}}

\newcommand{\vp}{\ensuremath{\text{vp}}}

\begin{document}

\title{Vacuum polarization in a one-dimensional effective quantum-electrodynamics model}
\author{Timoth\'ee Audinet}
\email{timothee.audinet@sorbonne-universite.fr}
\affiliation{Laboratoire de Chimie Th\'eorique, Sorbonne Universit\'e and CNRS, F-75005 Paris, France}
\author{Umberto Morellini}
\email{morellini@ceremade.dauphine.fr}
\affiliation{CEREMADE, Université Paris-Dauphine-PSL and CNRS, F-75016 Paris, France}
\affiliation{Laboratoire de Chimie Th\'eorique, Sorbonne Universit\'e and CNRS, F-75005 Paris, France}
\author{Antoine Levitt}
\email{antoine.levitt@universite-paris-saclay.fr}
\affiliation{Laboratoire de mathématiques d'Orsay, Université Paris-Saclay and CNRS, F-91405 Orsay, France}
\author{Julien Toulouse}
\email{toulouse@lct.jussieu.fr}
\affiliation{Laboratoire de Chimie Th\'eorique, Sorbonne Universit\'e and CNRS, F-75005 Paris, France}
\affiliation{Institut Universitaire de France, F-75005 Paris, France}

\date{March 9, 2025}

\begin{abstract}
With the aim of progressing toward a practical implementation of an effective quantum-electrodynamics (QED) theory of atoms and molecules, which includes the effects of vacuum polarization through the creation of virtual electron-positron pairs but without the explicit photon degrees of freedom, we study a one-dimensional effective QED model of the hydrogen-like atom with delta-potential interactions. This model resembles the three-dimensional effective QED theory with Coulomb interactions while being substantially simpler. We provide some mathematical details about the definition of this model, calculate the vacuum-polarization density, and the Lamb-type shift of the bound-state energy, correcting and extending results of previous works. We also study the approximation of the model in a finite plane-wave basis, and in particular we discuss the basis convergence of the bound-state energy and eigenfunction, of the vacuum-polarization density, and of the Lamb-type shift of the bound-state energy. We highlight the difficulty of converging the vacuum-polarization density in a finite basis and we propose a way to improve it. The present work could give hints on how to perform similar calculations for the three-dimensional effective QED theory of atoms and molecules.
\end{abstract}

\maketitle

\section{Introduction}

It is important to take into account the effects of special relativity in the quantum electronic-structure theory of atoms, molecules, and solids~\cite{Pyy-ARPC-12,Pyy-CR-12,Aut-JCP-12}. State-of-the-art relativistic electronic-structure calculations are based on the Dirac-Coulomb-Breit Hamiltonian in the no-pair approximation (see, e.g., Refs.~\onlinecite{SauVis-INC-03,DyaFae-BOOK-07,ReiWol-BOOK-09}). The next step is to go beyond the no-pair approximation, i.e. including the quantum-electrodynamics (QED) effect of virtual electron–positron pairs. This is not only important for highly accurate calculations, but also to put relativistic electronic-structure theory on deeper theoretical grounds. 

Highly accurate bound-state QED perturbative methods have been developed but are limited to few-electron atomic systems (see, e.g., Refs.~\onlinecite{MohPluSof-PR-98,Sha-PR-02,LinSalAse-PR-04,IndMoh-INC-17,NonMarMat-JCTC-24}). For many-electron atoms and molecules, it has been proposed to estimate QED corrections with model one-electron operators (see, e.g., Refs.~\onlinecite{PyyZha-JPB-03,ShaTupYer-PRA-13,SchPasPunBow-NPA-15,PasEliBorKalSch-PRL-17,MalGlaShaTupYerZay-PRA-22,Sal-THESIS-22,Skr-JCP-21,SunSalSau-JCP-22,FlyQuiGra-ARX-24a,FlyQuiGra-ARX-24b}). An appealing approach for ab initio relativistic electronic-structure calculations beyond the no-pair approximation is given by an effective QED theory, which includes the effects of vacuum polarization through the creation of virtual electron-positron pairs but still uses a static Coulomb or Coulomb-Breit two-particle interaction instead of explicit photons (see, e.g., Refs.~\cite{ChaIra-JPB-89,SauVis-INC-03,Kut-CP-12,LiuLin-JCP-13,Liu-PR-14,Liu-IJQC-15,Liu-JCP-20,Tou-SPC-21,Liu-WIRES-22,AudTou-JCP-23}). This effective QED theory with the Coulomb two-particle interaction has been the subject of a number of detailed mathematical studies which established the soundness of this approach at the Hartree-Fock level~\cite{HaiLewSer-CMP-05,HaiLewSer-JPA-05,HaiLewSol-CPAM-07,HaiLewSerSol-PRA-07,GraLewSer-CMP-09,HaiLewSer-ARMA-09,GraLewSer-CMP-11,Lew-INC-11}. Based on this effective QED theory, it has been proposed to formulate a relativistic density-functional theory~\cite{Tou-SPC-21} and a relativistic reduced density-matrix functional theory~\cite{RodGieVis-SPC-22}. 

As in full QED, the difficulty with this effective QED theory lies in the fact that it contains infinities. In particular, the vacuum-polarization density diverges in the ultraviolet (UV) limit (see, e.g., Ref.~\onlinecite{HaiSie-CMP-03}). This UV divergence can be regularized with a UV momentum cutoff and the dependence on the cutoff can be absorbed into a redefinition of the elementary charge, which is called charge renormalization (see, e.g., Refs.~\onlinecite{HaiLewSerSol-PRA-07,Lew-INC-11}). However, it is presently not clear how to deal with this situation in a finite basis and consequently no practical implementation of this effective QED theory has been done so far.

To progress toward the goal of a practical implementation of the above-mentioned effective QED theory for atomic and molecular calculations, this theory was studied in Ref.~\onlinecite{AudTou-JCP-23} in the context of a one-dimensional (1D) model of the relativistic hydrogen-like atom using delta-potential interactions. In the non-relativistic version of this model~\cite{Fro-JCP-56,HerSti-PRA-75,Her-JCP-86,TraGinTou-JCP-22}, the use of the delta potential is motivated by the fact that it leads to the same ground-state energy and wave function as the ground-state energy and radial wave function of the three-dimensional (3D) hydrogen-like atom with the Coulomb potential. The relativistic version of this model without QED effects was also previously studied~\cite{SubBha-JPC-72,Lap-AJP-83,FilLorBan-JPA-12,GuiMunPirSan-FP-19}. The calculation of the vacuum-polarization density in this model was first attempted in Ref.~\onlinecite{NogBea-EL-86}. A more thorough study of the QED effects, including the Lamb-type shift of the bound-state energy, was performed in Ref.~\onlinecite{AudTou-JCP-23}. The interest in this 1D effective QED model lies in the fact that it resembles the 3D effective QED theory with Coulomb interactions while being substantially simpler.

In the present work, we reexamine in more mathematical details this 1D effective QED model, correcting and extending results of previous works. In particular, we perform a more careful calculation of the exact vacuum-polarization density and show that there is a Dirac-delta contribution that was missed in the previous calculations. We also study the calculation of the vacuum-polarization density in a finite plane-wave basis, highlighting the difficulty of converging this quantity with the size of the basis, and we propose a way to improve the convergence of the calculation.  

The paper is organized as follows. In Section~\ref{sec:exact}, we provide the mathematical definition of the 1D hydrogen-like Dirac model, we perform the calculation of the exact vacuum-polarization density, and discuss the resulting QED correction to the bound-state energy. In Section~\ref{sec:finitedim}, we study the approximation of the model in a finite plane-wave basis: we discuss the convergence of the bound-state energy and eigenfunction, the convergence of the vacuum-polarization density, and the convergence of the QED correction to the bound-state energy with respect to the size of the basis. Finally, Section~\ref{sec:concl} contains our conclusions. In the appendices, we provide some mathematical details about the definition of the Hamiltonian of the present model and about the rate of convergence of the bound-state energy in a plane-wave basis.

\section{Exact 1D hydrogen-like Dirac model}
\label{sec:exact}

In this section, we study the exact 1D hydrogen-like Dirac model in the infinite-dimensional setting.

\subsection{Hamiltonian}
\label{sec:hamiltonian}

We consider a 1D spinless relativistic electron with two-component states in the Hilbert space $\calh  = L^2(\mathbb{R},\mathbb{C}^2)$. The 1D hydrogen-like Dirac Hamiltonian with an electrostatic-type nucleus-electron Dirac-delta potential can be formally defined as~\cite{Lap-AJP-83,AudTou-JCP-23}
\begin{eqnarray}
\bm{D}_Z = \bm{D}_0 - Z \delta(x) \b{I}_2,
\label{Dv}
\end{eqnarray}
where $\bm{D}_0$ is the 1D free $2\times2$ Dirac Hamiltonian
\begin{eqnarray}
\bm{D}_0 = c \bm{\sigma}_1 \; p_x + \bm{\sigma}_3 \; mc^2,
\label{freeDiracHam}
\end{eqnarray}
where $p_x= -\i \d/\d x$ is the momentum operator, $c$ is the speed of light, $m$ is the electron mass, and $\bm{\sigma}_1$ and $\bm{\sigma}_3$ are the $2\times2$ Pauli matrices
\begin{eqnarray}
\bm{\sigma}_1 = \left(\begin{array}{cc}
0&1\\
1&0\\
\end{array}\right)
~~\text{and}~~
\bm{\sigma}_3 = \left(\begin{array}{cc}
1&0\\
0&-1\\
\end{array}\right),
\end{eqnarray}
and $Z\geq 0$ is the nuclear charge and $\b{I}_2$ is the $2\times2$ identity matrix. For all the other usual physical constants, we always assume atomic units in which $\hbar = e = 4\pi \epsilon_0 = 1$.

The delta potential in Eq.~(\ref{Dv}) makes in fact the definition of $\bm{D}_Z$ ambiguous. There are several self-adjoint operators $\bm{D}_Z$ compatible with the above formal definition. As in Ref.~\onlinecite{AudTou-JCP-23}, we choose the self-adjoint operator $\bm{D}_Z$ defined as having the same action of $\bm{D}_0$ for $x\neq 0$, i.e.
\begin{eqnarray}
\bm{D}_Z \bm{\psi} = \bm{D}_0 \bm{\psi} \quad \text{on}\; \mathbb{R}\!\setminus\!\{0\},
\label{Dpsi=D0psi}
\end{eqnarray}
with the $Z$-dependent domain~\cite{Seb-LMP-89,BenDab-LMP-94,Hug-RMP-97,Hug-JMAA-99,PanRic-JMP-14}
\begin{eqnarray}
\text{Dom}(\bm{D}_Z) = \left\{ \bm{\psi} \in H^1(\mathbb{R}\! \setminus \! \{0\},\mathbb{C}^2) \; |\; \bm{\psi}(0^+) = \b{M}_Z \bm{\psi}(0^-) \right\}.
\label{DomD}
\end{eqnarray}
In Eq.~(\ref{DomD}), $H^1(\mathbb{R} \! \setminus \! \{0\},\mathbb{C}^2) \equiv H^1(\mathbb{R}^-,\mathbb{C}^2) \oplus H^1(\mathbb{R}^+,\mathbb{C}^2)$ is the first-order broken Sobolev space (allowing for a non-square-integrable derivative only at $x=0$) expressed with the standard first-order Sobolev space $H^1(\Omega,\mathbb{C}^2)=\{ \bm{\psi} \in L^2(\Omega,\mathbb{C}^2) \; |\; \d \bm{\psi}/\d x \in L^2(\Omega,\mathbb{C}^2) \}$ for a domain $\Omega \subseteq \mathbb{R}$, and $\b{M}_Z$ is the following unitary $2\times 2$ matrix enforcing a boundary condition at $x=0$~\cite{SubBha-JPC-72,Lap-AJP-83,FilLorBan-JPA-12}
\begin{eqnarray}
\b{M}_Z &=& \left(\begin{array}{cc}
\cos \theta    & \i \sin \theta\\
\i \sin \theta & \cos \theta\\
\end{array}\right),
\label{1stboundarycond}
\end{eqnarray}
with $\theta=2\arctan(Z/2c)$.

The Hamiltonian $\bm{D}_Z$ has a single bound state with energy~\cite{Lap-AJP-83,CouNog-PRA-87,NogBea-EL-86,AudTou-JCP-23}
\begin{eqnarray}
\varepsilon_\text{b}^Z  = mc^2 \frac{1-(Z/2c)^2}{1+(Z/2c)^2},
\label{epsilon0tilde}
\end{eqnarray}
and eigenfunction
\begin{eqnarray}
\bm{\psi}_\text{b}^Z(x) = A_\text{b}
\left(\begin{array}{c}
1\\
\i \sgn(x) Z/2c
\end{array}\right) e^{-\kappa_\text{b} |x|},
\label{psi0tilde}
\end{eqnarray}
where $\sgn$ is the sign function, $\kappa_\text{b} = m Z/(1+(Z/2c)^2)$, and $A_\text{b}=\sqrt{\kappa_\text{b}/(1+(Z/2c)^2)}$. Note that the large (upper) component of the bound-state eigenfunction is an even function of $x$, while the small (lower) component is an odd function of $x$ with a discontinuity at $x=0$. Beside the eigenvalue $\varepsilon_\text{b}^Z$, the Hamiltonian $\bm{D}_Z$ has also a continuous energy spectrum $(-\infty,-mc^2]\cup [mc^2,+\infty)$. At $Z=0$, only the continuous energy spectrum $(-\infty,-mc^2]\cup [mc^2,+\infty)$ remains. The bound-state energy in Eq.~(\ref{epsilon0tilde}) is strictly positive for $Z < 2 c$. Moreover, the bound-state energy level never dives into the negative-energy continuum for all $Z>0$. The bound-state energy only approaches the top of the negative-energy continuum as $Z\to\infty$, i.e.  $\lim_{Z\to \infty}\varepsilon_\text{b}^Z=-mc^2$. Hence, there are no supercritical QED effects in the present model~\cite{LoeSun-JPA-90,NogParToy-JPA-90}.

In Appendix~\ref{app:extension}, we argue that the matrix elements of the Hamiltonian $\bm{D}_Z$ can be defined on a larger $Z$-independent set of functions with the expression
\begin{eqnarray}
\braket{\bm{\phi}}{\bm{D}_Z \bm{\psi}} &=& \braket{\bm{\phi}}{\bm{D}_0 \bm{\psi}} - Z \bar{\bm{\phi}}^\dagger(0) \bar{\bm{\psi}}(0),
\label{psiDphi}
\end{eqnarray}
where $\braket{.\;}{.}$ designates the inner product of $L^2(\mathbb{R},\mathbb{C}^2)$, and $\bar{\bm{\psi}}(0) = [\bm{\psi}(0^+) + \bm{\psi}(0^-)]/2$ and $\bar{\bm{\phi}}(0) = [\bm{\phi}(0^+) + \bm{\phi}(0^-)]/2$. Importantly, Eq.~(\ref{psiDphi}) can be used to calculate the matrix elements of the Hamiltonian on a basis of functions which are continuous and thus which do not belong to the domain considered in Eq.~(\ref{DomD}).

\subsection{Vacuum-polarization density}
\label{sec:vpd}

The vacuum-polarization density $n^\vp(x)$ for the 1D hydrogen-like Dirac model was calculated in Refs.~\onlinecite{NogBea-EL-86,AudTou-JCP-23}. However, these calculations were only valid for $x\neq 0$. Here, we reexamine the calculation of the vacuum-polarization density using the momentum-space Green function and show that there is a Dirac-delta function contribution at $x=0$ that was missed in the previously cited works. In order to make things mathematically simpler, we will now work on the following Hilbert space with a UV momentum cutoff parameter $\Lambda$ (similarly to the 3D case, see e.g. Ref.~\onlinecite{HaiLewSerSol-PRA-07})
\begin{eqnarray}
\calh_\Lambda = \left\{ \bm{\psi} \in \calh \;|\; \hat{\bm{\psi}}(p)=\bm{0} \; \text{for}\; |p|>\Lambda \right\},
\label{}
\end{eqnarray}
where $\hat{\bm{\psi}}$ is the Fourier transform of $\bm{\psi}$. We will then be interested in the infinite UV momentum cutoff limit, i.e. $\Lambda \to \infty$.

\subsubsection{General expression in terms of the Green function}
\label{sec:GeneralGreen}

The formal definition of the Hamiltonian in Eq.~(\ref{Dv}) or the definition via matrix elements in Eq.~(\ref{psiDphi}) leads to the following expression for the 1D hydrogen-like Dirac Hamiltonian in momentum space (for $|p|\leq \Lambda$ and  $|p'|\leq \Lambda$)
\begin{eqnarray}
\bm{D}_Z(p,p') = \bm{D}_0(p,p') + \bm{V}(p,p'),
\label{}
\end{eqnarray}
with
\begin{eqnarray}
\bm{D}_0(p,p') = \delta(p-p') \left[ c \bm{\sigma}_1 p + \bm{\sigma}_3 m c^2 \right] \;\text{and}\;
 \bm{V}(p,p') = - \frac{Z}{2\pi}\b{I}_2.
\label{DOpppVppp}
\end{eqnarray}
The Green function (or resolvent) operator $\bm{G}_0(\omega)=(\omega \b{I}_2 - \bm{D}_0)^{-1}$ of the 1D free Dirac Hamiltonian $\bm{D}_0$ in momentum space is, for $\omega \in \mathbb{C} \setminus \sigma(\bm{D}_0)$ where $\sigma(\bm{D}_0)$ is the spectrum of $\bm{D}_0$,
\begin{eqnarray}
\bm{G}_0(p,p';\omega) = \frac{\delta(p-p')}{\omega^2-\varepsilon_p^2}
\left(\begin{array}{cc}
mc^2 + \omega & c p  \\
c p & -m c^2  +\omega
\end{array}\right),
\label{Gppomega}
\end{eqnarray}
with $\varepsilon_p = \sqrt{p^2 c^2 + m^2 c^4}$. The Green function operator $\bm{G}_Z(\omega)=(\omega \b{I}_2 - \bm{D}_Z)^{-1}$ of the 1D hydrogen-like Dirac Hamiltonian $\bm{D}_Z$ in momentum space satisfies the Dyson equation, for $\omega \in \mathbb{C} \setminus \sigma(\bm{D}_Z)$,
\begin{eqnarray}
\bm{G}_Z(p,p';\omega) = \bm{G}_0(p,p';\omega) + \int_{-\Lambda}^\Lambda \int_{-\Lambda}^\Lambda \bm{G}_0(p,p_1;\omega) \bm{V}(p_1,p_2) \bm{G}_Z(p_2,p';\omega) \d p_1 \d p_2.
\label{Dysoneq}
\end{eqnarray}
Using the expression of $\bm{V}(p,p')$, the Dyson equation can be simplified as
\begin{eqnarray}
\bm{G}_Z(p,p';\omega) = \bm{G}_0(p,p';\omega) - \frac{Z}{2\pi} \bar{\bm{G}}_0(p;\omega) \bar{\bm{G}}_Z(p';\omega),
\label{GpppomegaDyson}
\end{eqnarray}
where $\bar{\bm{G}}_0(p,\omega)  = \int_{-\Lambda}^\Lambda \bm{G}_0(p,p_1;\omega) \d p_1$ and $\bar{\bm{G}}_Z(p';\omega)=\int_{-\Lambda}^\Lambda \bm{G}_Z(p_2,p';\omega)  \d p_2$. Integrating Eq.~(\ref{GpppomegaDyson}) over $p$ gives
\begin{eqnarray}
\bar{\bm{G}}_Z(p';\omega) = \bar{\bm{G}}_0(p';\omega) - \frac{Z}{2\pi} \bar{\bar{\bm{G}}}_0(\omega) \bar{\bm{G}}_Z(p';\omega),
\label{}
\end{eqnarray}
where $\bar{\bar{\bm{G}}}_0(\omega)=\int_{-\Lambda}^\Lambda \bar{\bm{G}}_0(p;\omega)  \d p$, and thus
\begin{eqnarray}
\bar{\bm{G}}_Z(p';\omega) = \left[ \b{I}_2 + \frac{Z}{2\pi} \bar{\bar{\bm{G}}}_0(\omega) \right]^{-1}\bar{\bm{G}}_0(p';\omega).
\label{}
\end{eqnarray}
Inserting the last expression in Eq.~(\ref{GpppomegaDyson}), we obtain for the variation of the Green function due to the nucleus-electron potential, $\Delta \bm{G}_Z(p,p';\omega) = \bm{G}_Z(p,p';\omega) - \bm{G}_0(p,p';\omega)$,
\begin{eqnarray}
\Delta \bm{G}_Z(p,p';\omega) = - \frac{Z}{2\pi} \bar{\bm{G}}_0(p;\omega) \left[ \b{I}_2 + \frac{Z}{2\pi} \bar{\bar{\bm{G}}}_0(\omega) \right]^{-1}\bar{\bm{G}}_0(p';\omega).
\label{DGZpppw}
\end{eqnarray}
From Eq.~(\ref{Gppomega}), we can calculate $\bar{\bm{G}}_0(p;\omega)$ and $\bar{\bar{\bm{G}}}_0(\omega)$,
\begin{eqnarray}
\bar{\bm{G}}_0(p;\omega)= \frac{1}{\omega^2-\varepsilon_p^2}
\left(\begin{array}{cc}
mc^2 + \omega & c p  \\
c p & -m c^2  +\omega
\end{array}\right),
\label{barG0pomega}
\end{eqnarray}
and
\begin{eqnarray}
\bar{\bar{\bm{G}}}_0(\omega)
= \frac{\pi \; \xi(\Lambda,\omega)}{c} \left(\begin{array}{cc}
-g(\omega) & 0  \\
0 & g(-\omega)
\end{array}\right),
\label{barbarG0omega}
\end{eqnarray}
with $g(\omega) = \sqrt{(mc^2 + \omega)/(mc^2-\omega)}$ and $\xi(\Lambda,\omega) = (2/\pi) \arctan (c \Lambda/\sqrt{m^2 c^4 - \omega^2})$. Note that the function $\xi(\Lambda,\omega)$ reduces to $1$ in the infinite UV momentum cutoff limit, i.e. $\lim_{\Lambda \to \infty} \xi(\Lambda,\omega) = 1$. Note also that the zero off-diagonal elements in Eq.~(\ref{barbarG0omega}) come from the fact that we have integrated an odd function of $p$ over the symmetric interval $[-\Lambda,\Lambda]$, i.e. $\int_{-\Lambda}^\Lambda  cp/(\omega^2-\varepsilon_p^2) \d p = 0$. A non-symmetric UV momentum cutoff gives non-zero off-diagonal elements (see Appendix~\ref{app:asymmcutoff}). So, the Green function obtained in the limit $\Lambda\to\infty$ depends on how the UV momentum cutoff is chosen. The symmetric UV momentum cutoff correctly gives, in the limit $\Lambda\to\infty$, the Green function corresponding to the Hamiltonian defined by Eqs.~(\ref{Dpsi=D0psi}) and~(\ref{DomD}), which was already calculated in position space using different methods in Refs.~\onlinecite{Seb-LMP-89,AudTou-JCP-23}.
From Eqs.~(\ref{DGZpppw})-(\ref{barbarG0omega}), we finally arrive at the expression of the variation of the Green function in momentum space
\begin{eqnarray}
\Delta \bm{G}_Z(p,p';\omega) =  - \frac{Z}{2\pi} \frac{z_1(\omega) \b{A}_1(p,p';\omega)+ z_2(\omega) \b{A}_2(p,p';\omega)}{(\omega^2-\varepsilon_p^2)(\omega^2-\varepsilon_{p'}^2)},
\label{}
\end{eqnarray}
with $z_1(\omega) = (1-Z g(\omega)\xi(\Lambda,\omega)/2c)^{-1}$ and $z_2(\omega)=(1+Z g(-\omega)\xi(\Lambda,\omega)/2c)^{-1}$, and $\b{A}_1(p,p';\omega)$ and $\b{A}_2(p,p';\omega)$ are the following matrices
\begin{eqnarray}
\b{A}_1(p,p';\omega) = \left(\begin{array}{cc}
(mc^2+\omega)^2 & c p' (mc^2 +\omega)  \\
c p (mc^2+\omega) & c^2 p p'
\end{array}\right)
\;\text{and}\;
\b{A}_2(p,p';\omega) = \left(\begin{array}{cc}
c^2 p p' & c p (-mc^2 +\omega)  \\
c p' (-mc^2+\omega) & (-mc^2+\omega)^2
\end{array}\right).
\label{}
\end{eqnarray}

The Fourier transform of the vacuum-polarization density matrix is given by (see, e.g., Refs.~\onlinecite{WicKro-PR-56,HaiSie-CMP-03,Hai-AHP-04,HaiLewSerSol-PRA-07})
\begin{eqnarray}
\hat{\b{n}}_1^{\text{vp}}(p,p') = \frac{1}{2\pi} \int_{-\infty}^\infty \Delta \bm{G}_Z(p,p';\i u + \gamma) \d u,
\label{n1vppp}
\end{eqnarray}
where $\gamma$ is a real constant such that $-mc^2 < \gamma < \varepsilon_\text{b}^Z$, and the Fourier transform of the vacuum-polarization local density matrix can then be obtained as (see, e.g., Refs.~\onlinecite{HaiSie-CMP-03,EstLewSer-BAMS-08,HaiLewSer-ARMA-09}) 
\begin{eqnarray}
\hat{\b{n}}^{\text{vp}}_\Lambda(k) &=&  \frac{1}{\sqrt{2\pi}} \int_{\substack{|p+k/2|\leq \Lambda\\ |p-k/2|\leq \Lambda}} \hat{\b{n}}_1^{\text{vp}} (p+k/2,p-k/2) \d p,
\label{nvpkint}
\end{eqnarray}
where, for clarity, we have explicitly indicated the dependence on the UV momentum cutoff parameter $\Lambda$ in $\hat{\b{n}}^{\text{vp}}_\Lambda(k)$.
The constant $\gamma$ in Eq.~(\ref{n1vppp}) is there so that the integration over $u$ selects only the continuous negative-energy spectrum and not the bound-state energy $\varepsilon_\text{b}^Z$. From now on, we will assume that $Z <2 c$ so that the bound-state eigenvalue $\varepsilon_\text{b}^Z$ is always strictly positive, and we will choose $\gamma=0$. What we call the vacuum-polarization local density matrix is thus obtained as the inverse Fourier transform of $\hat{\b{n}}^{\text{vp}}_\Lambda(k)$
\begin{eqnarray}
\b{n}^{\text{vp}}_\Lambda(x) &=&  \frac{1}{\sqrt{2\pi}} \int_{-\infty}^\infty \hat{\b{n}}^{\text{vp}}_\Lambda(k) e^{\i k x}\d k,
\label{nvpxfromFT}
\end{eqnarray}
i.e., it is the position-space diagonal of the vacuum-polarization density matrix. 
Note that the slowest-decaying terms in the integrand in Eq.~(\ref{n1vppp}) decay as $1/u^2$, so that the corresponding integral over $u$ is convergent. Note also that, due to the integration domain in Eq.~(\ref{nvpkint}), $\hat{\b{n}}^{\text{vp}}_\Lambda(k)$ is zero for $|k|> 2\Lambda$. Hence, the integral in Eq.~(\ref{nvpxfromFT}) is convergent. 

We will see below that, in the limit $\Lambda\to\infty$, the Fourier transform $\hat{\b{n}}^{\text{vp}}_\Lambda(k)$ in Eq.~(\ref{nvpkint}) contains terms that go to a constant as $k\to \pm\infty$ and thus, in this limit, the integral in Eq.~(\ref{nvpxfromFT}) contains a Dirac-delta contribution. In fact, in the  limit $\Lambda\to\infty$, the density matrix is not a trace-class operator and defining its associated density is not a priori obvious.

\subsubsection{Uehling vacuum-polarization density}

The integral in Eq.~(\ref{n1vppp}) can be analytically calculated at first order in $Z$, which gives the Fourier transform of the 1D analog of the Uehling vacuum-polarization density matrix
\begin{eqnarray}
\hat{\b{n}}_1^{\text{vp},(1)}(p,p') &=& \frac{1}{2\pi} \int_{-\infty}^\infty \Delta \bm{G}^{(1)}_Z(p,p';\i u) \d u
\nonumber\\
                                   &=& -\frac{Z}{4\pi(\varepsilon_{p}^2 \varepsilon_{p'} + \varepsilon_{p} \varepsilon_{p'}^2 )}
\left(\begin{array}{cc}
m^2 c^4 + c^2 p p' - \varepsilon_{p}\varepsilon_{p'} & - m c^3  (p -p')  \\
m c^3  (p -p')  & m^2 c^4 + c^2 p p' - \varepsilon_{p}\varepsilon_{p'}
\end{array}\right),
\label{n1vpUpp}
\end{eqnarray}
which is independent of $\Lambda$, and the Fourier transform of the Uehling vacuum-polarization local density matrix is then obtained using Eq.~(\ref{nvpkint})
\begin{eqnarray}
\hat{\b{n}}^{\text{vp},(1)}_\Lambda(k) &=&  \frac{1}{\sqrt{2\pi}}  \int_{\substack{|p+k/2|\leq \Lambda\\ |p-k/2|\leq \Lambda}} \hat{\b{n}}_1^{\text{vp},(1)} (p+k/2,p-k/2) \d p.
\label{nvpUkmatLambda}
\end{eqnarray}
The expression of $\hat{\b{n}}^{\text{vp},(1)}_\Lambda(k)$ resulting from the integration in Eq.~(\ref{nvpUkmatLambda}) is lengthy, but it has a relatively simple expression in the infinite UV momentum cutoff limit,
\begin{eqnarray}
\hat{\b{n}}^{\text{vp},(1)}(k) = \lim_{\Lambda \to \infty}\hat{\b{n}}^{\text{vp},(1)}_\Lambda(k) &=&  \frac{{\cal N}_0^{\text{vp},(1)}}{2\sqrt{2\pi}} \b{I}_2 + \hat{\b{n}}^{\text{vp},(1)}_\text{reg}(k),
\label{nvpUkmat}
\end{eqnarray}
which contains a diagonal constant contribution with ${\cal N}_0^{\text{vp},(1)} = Z/(\pi c)$ and a regular contribution which goes to zero as $k \to \pm \infty$
\begin{eqnarray}
\hat{\b{n}}^{\text{vp},(1)}_\text{reg}(k)  &=& -\frac{Zm}{(2\pi)^{3/2} h_k} 
\left(\begin{array}{cc}
\dfrac{4 m c}{k} \arctanh\left(\dfrac{k}{h_k}\right) & \ln\left( \dfrac{h_k-k}{h_k+k} \right)  \\
-\ln\left( \dfrac{h_k-k}{h_k+k} \right) & \dfrac{4 m c}{k} \arctanh\left(\dfrac{k}{h_k}\right)
\end{array}\right),
\end{eqnarray}
with $h_k = \sqrt{k^2 + 4 m^2 c^2}$. Note that, in contrast with the 3D case (see, e.g., Refs.~\onlinecite{HaiSie-CMP-03,HaiLewSerSol-PRA-07}), here there is no divergence in the infinite UV momentum cutoff limit. Therefore, we will implicitly consider this limit in the rest of Section~\ref{sec:exact}. Taking the inverse Fourier transform in Eq.~(\ref{nvpUkmat}), we find the Uehling vacuum-polarization local density matrix in position space (in the distribution sense)
\begin{eqnarray}
\b{n}^{\text{vp},(1)}(x) &=&  \frac{{\cal N}_0^{\text{vp},(1)}}{2} \delta(x) \b{I}_2 + \b{n}^{\text{vp},(1)}_\text{reg}(x),
\label{nvpUxmat}
\end{eqnarray}
where the regular contribution can be expressed as, for $x\not=0$, (see Appendix D of Ref.~\onlinecite{AudTou-JCP-23})
\begin{eqnarray}
\b{n}^{\text{vp},(1)}_\text{reg}(x) = -\frac{Z}{4 c^2} \int_{-\infty}^{\infty} \frac{\d u}{2\pi} \; e^{-2 \kappa(\i u)|x|} 
\left(\begin{array}{cc}
g(\i u)^2 +1 & -\i \sgn(x) [g(\i u)+ g(-\i u)] \\
\i \sgn(x) [g(\i u)+ g(-\i u)]  & g(-\i u)^2 +1
\end{array}\right),
\label{}
\end{eqnarray}
with $\kappa(\i u) = \sqrt{m^2 c^4 + u^2}/c$.
Finally, the Uehling vacuum-polarization density $n^{\text{vp},(1)}(x) = \tr[\b{n}^{\text{vp},(1)}(x)]$ (where $\tr$ designates the trace of a $2\times2$ matrix) has the expression
\begin{eqnarray}
n^{\text{vp},(1)}(x) = {\cal N}_0^{\text{vp},(1)} \delta(x) + n^{\text{vp},(1)}_\text{reg}(x),
\label{nvp1x}
\end{eqnarray}
where the regular contribution was obtained in Ref.~\onlinecite{AudTou-JCP-23} in the compact form
\begin{eqnarray}
n^{\text{vp},(1)}_\text{reg}(x) &=&  -\frac{Z m}{\pi} \int_{1}^{\infty} \frac{e^{-2 mc |x| t}}{t\sqrt{t^2-1}} \d t.
\label{}
\end{eqnarray}
Note that $n^{\text{vp},(1)}(x)$ represents an opposite charge density (or electron–excess density). The vacuum-polarization \textit{charge} density is the opposite: $\rho^{\text{vp},(1)}(x) = -n^{\text{vp},(1)}(x)$.

\begin{figure}
        \centering
        \includegraphics[width=0.40\textwidth,angle=-90]{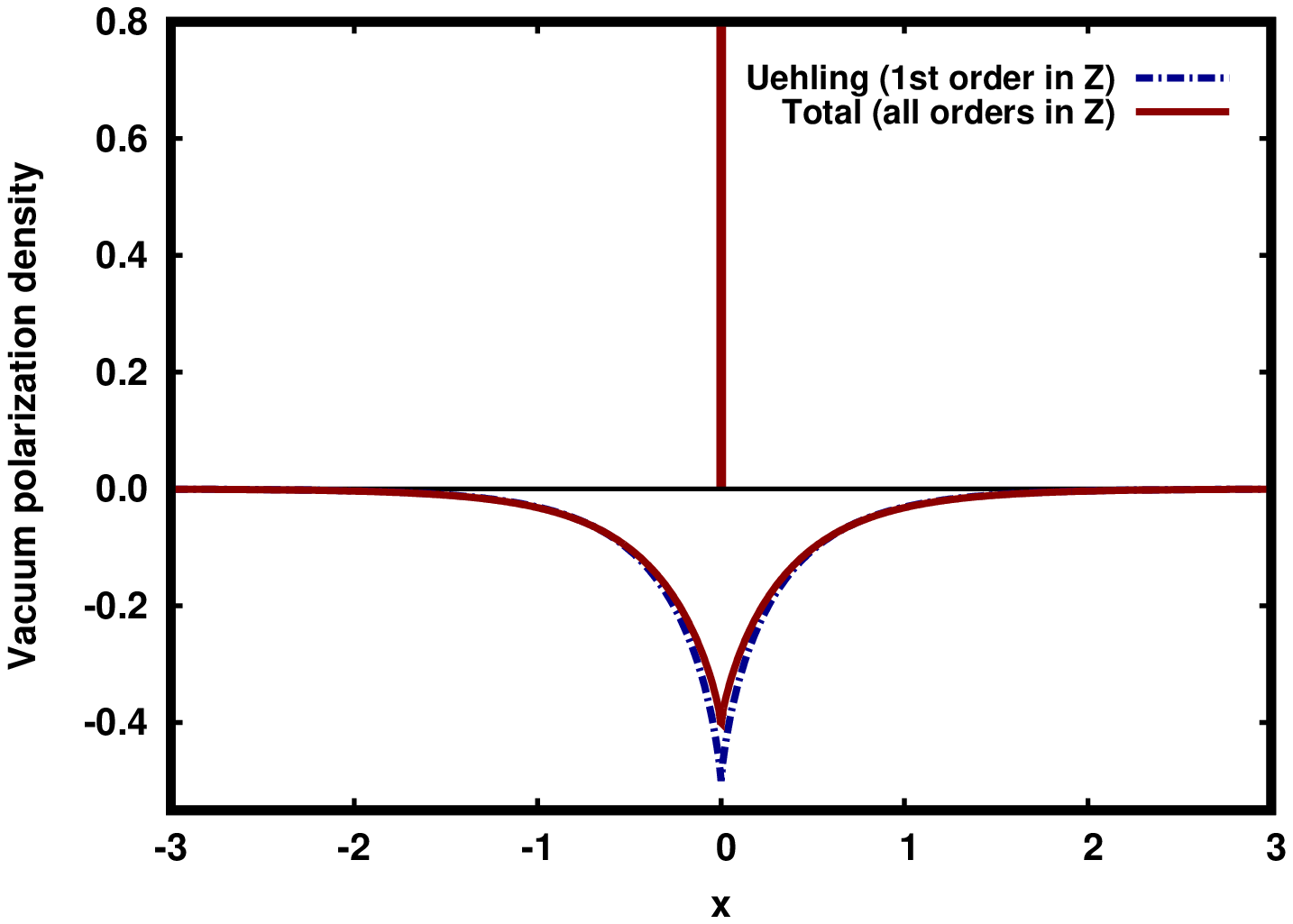}
        \caption{The Uehling and total vacuum-polarization densities $n^{\text{vp},(1)}(x)$ [Eq.~(\ref{nvp1x})] and $n^{\vp}(x)$ [Eq.~(\ref{nvpx})] for $m = c = Z = 1$. The vertical line represents a Dirac-delta function.}
        \label{fig:VPD}
\end{figure}

The Dirac-delta contribution in Eq.~(\ref{nvp1x}) was missed in Ref.~\onlinecite{AudTou-JCP-23}, in which the calculation was done entirely in position space. The present calculation which starts in momentum space is better suited to correctly catch the Dirac-delta contribution (see, however, Appendix \ref{app:error} for how to find the Dirac-delta contribution in position space). The coefficient in front of the Dirac-delta function is exactly the opposite of the spatial integral of the regular contribution $n^{\text{vp},(1)}_\text{reg}(x)$, so that the total Uehling vacuum-polarization density integrates to zero, i.e.
\begin{eqnarray}
\int_{-\infty}^\infty n^{\text{vp},(1)}(x) \d x = 0.
\label{}
\end{eqnarray}
The overall shape of the present Uehling vacuum-polarization density, represented in Fig.~\ref{fig:VPD}, is consistent with its usual interpretation as the spatial distribution of virtual electron-positron pairs induced by the nuclear charge (see, e.g., Ref.~\onlinecite{Sak-BOOK-67}), i.e. there is an excess of electrons at the nucleus (the Dirac-delta contribution) and an excess of positrons distributed farther away (the regular contribution). 

We emphasize that, in the  present 1D model, the Uehling vacuum-polarization density can be obtained without performing any renormalization. This is in sharp contrast with the case of the 3D hydrogen-like atom for which there is a contribution to the first-order vacuum-polarization density that diverges in momentum space and which is eliminated by charge renormalization. The Uehling vacuum-polarization density is then only the remaining contribution after renormalization (see, e.g., Refs.~\onlinecite{HaiSie-CMP-03,GreRei-BOOK-09}). This remaining 3D Uehling vacuum-polarization density integrates to zero~\cite{GreRei-BOOK-09,Hai-AHP-04}, as in the present 1D model. However, in the 3D hydrogen-like atom, the Uehling vacuum-polarization density (after charge renormalization) has a shape which is the opposite of the one obtained in the present 1D model, i.e. there is an excess of positrons at the nucleus and an excess of electrons farther away~\cite{GreRei-BOOK-09}. This shape of the Uehling vacuum-polarization density in the 3D hydrogen-like atom may seem counterintuitive but one has to keep in mind that it comes after charge renormalization which has eliminated an (infinite) excess of electrons at the nucleus.

\subsubsection{Total vacuum-polarization density}
\label{sec:totalvpd}

The case of the total vacuum-polarization density (i.e., at all orders in $Z$) is more complicated to study since we have not managed to analytically perform the integral in Eq.~(\ref{n1vppp}). Nevertheless, by a mixed analytical-numerical study, we have been able to determine that the Fourier transform of the total vacuum-polarization local density matrix has the form (still in the infinite UV momentum cutoff limit)
\begin{eqnarray}
\hat{\b{n}}^{\text{vp}}(k) = \frac{{\cal N}_0^\text{vp}}{2\sqrt{2\pi}} \b{I}_2 + \hat{\b{n}}^{\text{vp}}_\text{reg}(k),
\end{eqnarray}
with a diagonal constant contribution involving the quantity
\begin{eqnarray}
{\cal N}_0^\text{vp} = \frac{Z/c}{\pi(1+(Z/2c)^2)},
\label{}
\end{eqnarray}
and a regular contribution (which goes to zero as $k\to \pm \infty$) which, in practice, can be obtained by inverting the orders of the integrals in Eqs.~(\ref{n1vppp}) and~(\ref{nvpkint}), leading to
\begin{eqnarray}
\hat{\b{n}}^{\text{vp}}_\text{reg}(k) = -\frac{Z}{\sqrt{2\pi} c^2} \int_{-\infty}^{\infty} \frac{\d u}{2\pi} \frac{z_1(\i u) \b{B}_1(k,u) + z_2(\i u) \b{B}_2(k,u)}{\kappa(\i u) (k^2c^2 + 4 m^2 c^4 + 4 u^2)},
\label{}
\end{eqnarray}
with the matrices
\begin{eqnarray}
\b{B}_1(k,u) = \left(\begin{array}{cc}
(mc^2+\i u)^2 & - k c (m c^2 + \i u)/2\\
k c (m c^2+\i u)/2  & m^2c^4 + u^2
\end{array}\right)
\; \text{and} \; 
\b{B}_2(k,u) =
\left(\begin{array}{cc}
m^2 c^4 +u^2 & - k c (m c^2 - \i u)/2 \\
k c (m c^2-\i u)/2  & (mc^2 -\i u)^2
\end{array}\right).
\label{}
\end{eqnarray}
In position space, the total vacuum-polarization local density matrix is
\begin{eqnarray}
\b{n}^{\text{vp}}(x) =  \frac{{\cal N}_0^\text{vp}}{2} \delta(x) \b{I}_2 + \b{n}^{\text{vp}}_\text{reg}(x),
\label{n1vpx}
\end{eqnarray}
where the regular contribution can be expressed as, for $x\not=0$, (see Appendix D of Ref.~\onlinecite{AudTou-JCP-23})~\cite{AudMorLevTou-JJJ-XX-note1}
\begin{eqnarray}
\b{n}^{\text{vp}}_\text{reg}(x) = -\frac{Z}{4 c^2} \int_{-\infty}^{\infty} \frac{\d u}{2\pi} \; e^{-2 \kappa(\i u)|x|}  \left[ z_1(\i u) \b{C}_1(x,u) + z_2(\i u) \b{C}_2(x,u) \right],
\label{}
\end{eqnarray}
with the matrices
\begin{eqnarray}
\b{C}_1(x,u) = \left(\begin{array}{cc}
g(\i u)^2  & -\i \sgn(x) g(\i u)\\
\i \sgn(x) g(\i u)  & 1
\end{array}\right)
\; \text{and} \; 
\b{C}_2(x,u) =
\left(\begin{array}{cc}
1  & -\i \sgn(x) g(-\i u) \\
\i \sgn(x) g(-\i u)  & g(-\i u)^2
\end{array}\right).
\label{}
\end{eqnarray}
Finally, the total vacuum-polarization density $n^{\text{vp}}(x) = \tr[\b{n}^{\text{vp}}(x)]$ has the form
\begin{eqnarray}
n^{\text{vp}}(x) =  {\cal N}_0^\text{vp} \delta(x)  + n^{\text{vp}}_\text{reg}(x),
\label{nvpx}
\end{eqnarray}
where the regular part $n^{\text{vp}}_\text{reg}(x) =  \tr[\b{n}^{\text{vp}}_\text{reg}(x)]$ was given in different forms in Refs.~\onlinecite{NogBea-EL-86,AudTou-JCP-23}. Again, the reason why the Dirac-delta contribution was missed in Refs.~\onlinecite{NogBea-EL-86,AudTou-JCP-23} is that these references focused on pointwise calculations of the vacuum-polarization density in position space. The present momentum-space approach avoids this limitation.

The form of the total vacuum-polarization density is very similar to the one of the Uehling vacuum-polarization density, as shown in Fig.~\ref{fig:VPD}. However, an important difference is that, for $Z\neq 0$, the total vacuum-polarization density does not integrate to zero
\begin{eqnarray}
{\cal N}^\text{vp} = \int_{-\infty}^\infty n^{\text{vp}}(x) \d x  = {\cal N}_0^\text{vp} + {\cal N}_\text{reg}^\text{vp} \neq 0,
\label{}
\end{eqnarray}
where the integral of the regular contribution is~\cite{NogBea-EL-86,AudTou-JCP-23}
\begin{eqnarray}
{\cal N}_\text{reg}^\text{vp} = \int_{-\infty}^\infty n^{\text{vp}}_\text{reg}(x) \d x = -\frac{2}{\pi} \arctan\left(\frac{Z}{2c}\right).
\label{Nvpreg}
\end{eqnarray}
We may define the effective nuclear charge observed at a distance $d$ from the nucleus as
\begin{eqnarray}
Z_\text{obs} (d) = Z - \int_{-d}^{d} n^{\text{vp}}(x) \d x,
\label{}
\end{eqnarray}
with the limits 
\begin{eqnarray}
\lim_{d\to 0^+} Z_\text{obs} (d) = Z - {\cal N}_0^\text{vp} \leq Z,
\label{}
\end{eqnarray}
and 
\begin{eqnarray}
\lim_{d\to \infty} Z_\text{obs} (d) = Z - {\cal N}^\text{vp} \geq Z.
\label{}
\end{eqnarray}
Thus, at short distances $d$ the vacuum-polarization density screens the nuclear charge, resulting in an observed nuclear charge $Z_\text{obs}(d)$ which is smaller than the bare nuclear charge $Z$. At large distances $d$ from the nucleus ($d \gg \lambdabar$ where $\lambdabar = 1/(mc)$ is the reduced Compton wavelength providing a measure of the spatial extension of the vacuum-polarization density), the observed nuclear charge $Z_\text{obs}(d)$ is larger than the bare nuclear charge $Z$.

It may be tempting to think that the fact that the total vacuum-polarization density does not integrate to zero should lead to a (finite) charge renormalization. However, the present situation is different from the charge renormalization performed in the 3D case with a finite UV momentum cutoff (see, e.g., Ref.~\onlinecite{HaiLewSerSol-PRA-07}). Indeed, we believe that the fact that ${\cal N}^\text{vp}$ is not zero (and not even an integer) in the present model should be understood as an example of fermion-number fractionalization, which is a phenomenon known to appear for 1D Dirac equations with soliton-type potentials (see, e.g., Refs.~\onlinecite{GolWil-PRL-81,NieSem-PR-86,KriRoz-SPU-87,RaoSahPan-R-08}). The non-zero value of the fermion number is related to the infrared limit~\cite{KivSch-PRB-82,RajBel-PL-82,Nog-ARX-08}. Note, in particular, that the expression of ${\cal N}_\text{reg}^\text{vp}$ in Eq.~(\ref{Nvpreg}) has the general form of the fermion number found in soliton models~\cite{GolWil-PRL-81,NieSem-PR-86}. Also, we may note that, in the present model, for $Z=2c$ the (unoccupied) bound state has zero energy and the spectrum of $\bm{D}_Z$ is charge-conjugation symmetric, and ${\cal N}_\text{reg}^\text{vp}$ reduces to $-1/2$, which resembles the soliton scenario of Ref.~\onlinecite{JacReb-PRD-76} (see, also, Ref.~\onlinecite{RaoSahPan-R-08}).

In Ref.~\onlinecite{AudTou-JCP-23} (Section III.D and Appendix E), it was attempted to calculate the charge of the vacuum of the present model using the notion of ``$P^0$-trace'' of Refs.~\onlinecite{HaiLewSer-CMP-05,HaiLewSerSol-PRA-07,GraLewSer-CMP-09}. A rough numerical integration suggested that the $P^0$-trace formula of the charge of the vacuum gives zero, similarly to what is obtained in the 3D case in the presence of an UV momentum cutoff. However, for the present work, we have performed a more careful numerical integration which shows in fact that the $P^0$-trace formula of the (opposite) charge of the vacuum is numerically identical to the integral of the vacuum-polarization density, i.e. the fractional fermion number ${\cal N}^\text{vp}$. 
The fact that the $P^0$-trace formula of the charge of the vacuum is not an integer implies that the vacuum-polarization density matrix $\hat{\b{n}}_1^{\text{vp}}(p,p')$ [Eq.~(\ref{n1vppp})] of the present model is not a Hilbert-Schmidt operator and/or is not the difference of two projectors in the infinite UV momentum cutoff limit (see Refs.~\onlinecite{HaiLewSer-CMP-05,HaiLewSerSol-PRA-07}). If $\hat{\b{n}}_1^{\text{vp}}(p,p')$ were not a Hilbert-Schmidt operator, this would imply, according to the Shale-Stinespring theorem~\cite{ShaSti-JMM-65,Tha-BOOK-92}, that the polarized vacuum state of the present model, in the infinite UV momentum cutoff limit, could not be reached from a unitary transformation of the free vacuum state in the second-quantized Fock space (see, also, Refs.~\onlinecite{FalGro-LMP-87,GroOpe-NPB-87}). In any case, this confirms that the fact that the total vacuum-polarization density does not integrate to zero in the present 1D model is completely different from the situation of charge renormalization in 3D in the presence of an UV momentum cutoff. In the latter case, the vacuum-polarization density also does not integrate to zero but the vacuum-polarization density matrix is a Hilbert-Schmidt operator, the $P^0$-trace is an integer defining the charge of the vacuum, and the Shale-Stinespring theorem applies.

\begin{figure*}
        \centering
        \includegraphics[width=0.32\textwidth,angle=-90]{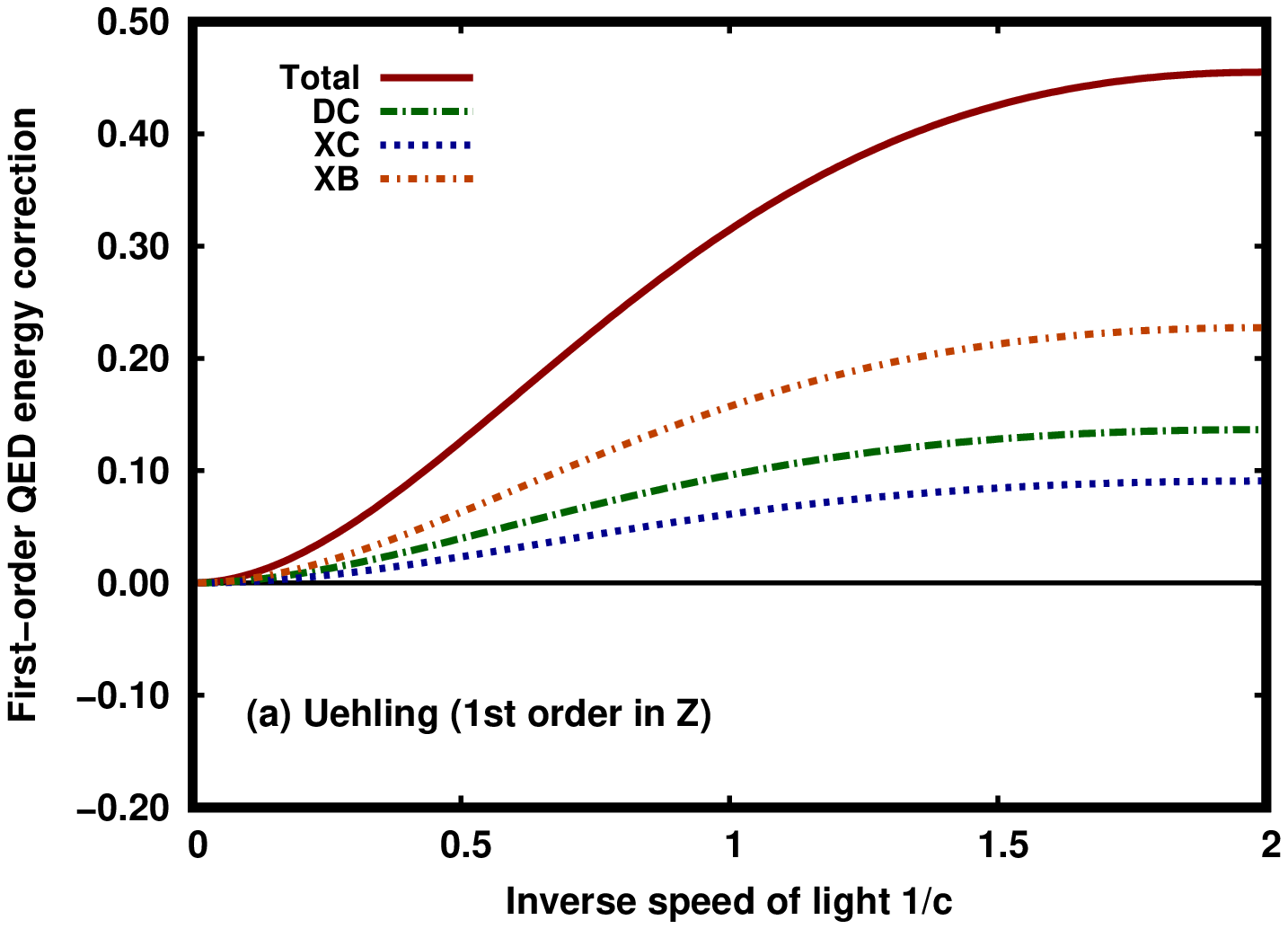}
        \includegraphics[width=0.32\textwidth,angle=-90]{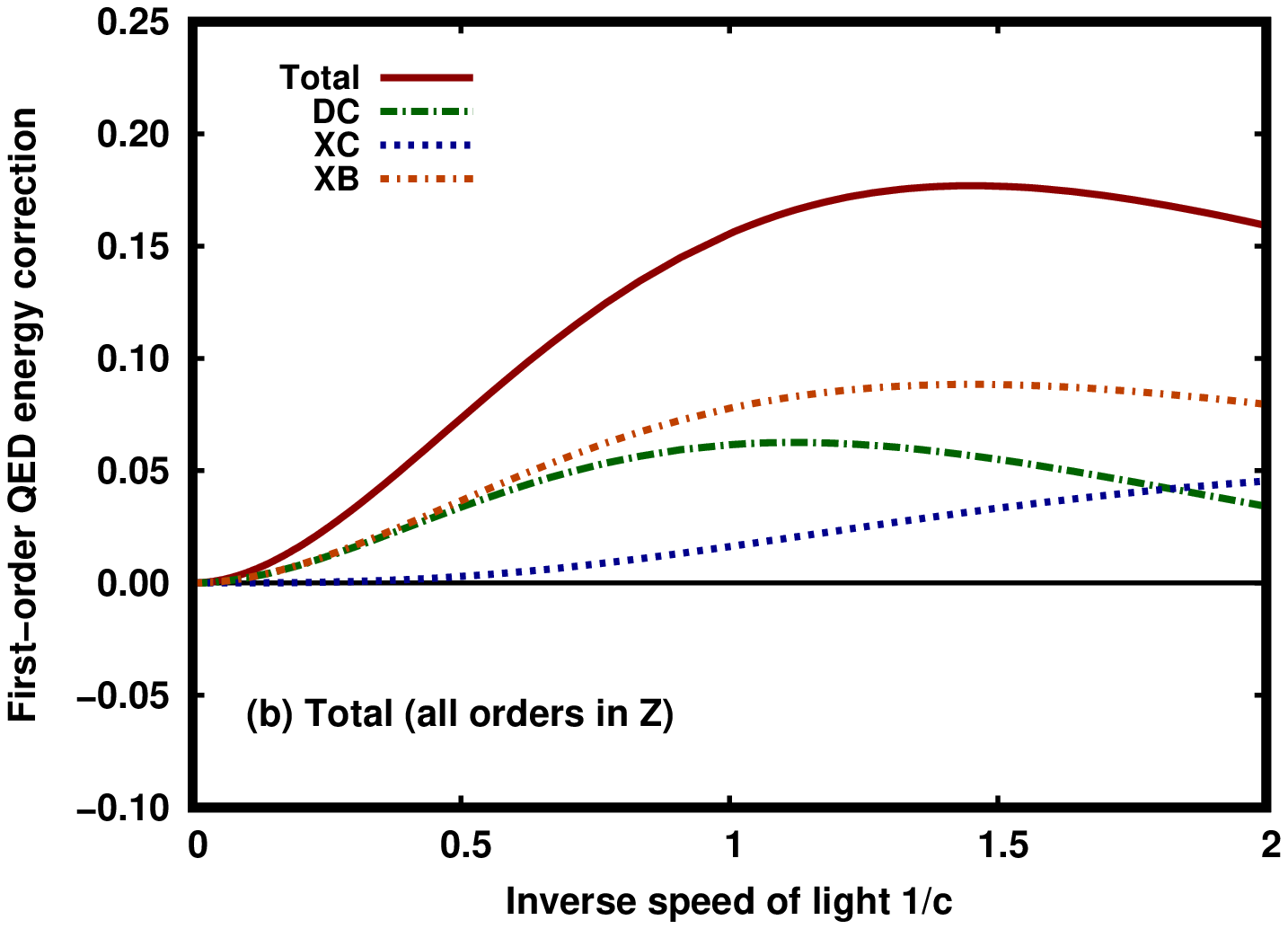}
        \caption{First-order QED correction to the bound-state energy [Eq.~(\ref{Evp1b})] for $m=Z=1$ as a function of the inverse speed of light $1/c$, using either the (a) Uehling or (b) total vacuum-polarization local density matrix [Eqs.~(\ref{nvpUxmat}) and (\ref{n1vpx})]. The total correction, as well as the direct-Coulomb-type (DC) [Eq.~\ref{Evp1bDC}], exchange-Coulomb-type (XC) [Eq.~\ref{Evp1bXC}], and exchange-Breit-type (XB) [Eq.~\ref{Evp1bXB}] contributions are shown. 
}
        \label{fig:lambshift}
\end{figure*}

\subsection{QED corrections to the bound-state energy}
\label{sec:QEDcorrections}

In the present 1D hydrogen-like Dirac model, we can calculate the shift of the bound-state energy due to the phenomenon of vacuum polarization, similarly to the Lamb shift in standard QED (see, e.g., Refs.~\onlinecite{MohPluSof-PR-98,SunSalSau-JCP-22}). The analog of the 3D Coulomb-Breit two-particle interaction in the present model is~\cite{CouNog-PRA-87,AudTou-JCP-23}
\begin{eqnarray}
\b{w}(x_1,x_2) = \delta(x_1-x_2) \left( \b{I}_2\otimes\b{I}_2 - \bm{\sigma}_1 \otimes \bm{\sigma}_1 \right),
\label{wx1x2}
\end{eqnarray}
where the first and second terms are the 1D analogs of the Coulomb and Breit interactions, respectively (see, e.g., Refs.~\onlinecite{DyaFae-BOOK-07,ReiWol-BOOK-09,Tou-SPC-21}). Note that, in 1D, the Breit interaction exactly reduces to the magnetic Gaunt interaction~\cite{AudTou-JCP-23}. At first order with respect to the two-particle interaction, the QED correction to the bound-state energy contains four contributions~\cite{AudTou-JCP-23}
\begin{eqnarray}
{\cal E}^{\text{vp},(1)}_\text{b} = {\cal E}^{\text{vp},(1),\text{DC}}_\text{b} + {\cal E}^{\text{vp},(1),\text{XC}}_\text{b} + {\cal E}^{\text{vp},(1),\text{DB}}_\text{b} + {\cal E}^{\text{vp},(1),\text{XB}}_\text{b}.
\label{Evp1b}
\end{eqnarray}
The direct-Coulomb-type (DC) and exchange-Coulomb-type (XC) contributions are
\begin{eqnarray}
{\cal E}^{\text{vp},(1),\text{DC}}_\text{b} = \int_{-\infty}^\infty n^\text{el}(x) n^\text{vp}(x) \d x,
\label{Evp1bDC}
\end{eqnarray}
and
\begin{eqnarray}
{\cal E}^{\text{vp},(1),\text{XC}}_\text{b} = -\int_{-\infty}^\infty \tr[ \b{n}^\text{el}(x) \b{n}^\text{vp}(x) ] \d x,
\label{Evp1bXC}
\end{eqnarray}
where $\b{n}^\text{el}(x) = \bm{\psi}_\text{b}(x) \bm{\psi}^\dagger_\text{b}(x)$ is the electronic bound-state local density matrix and $n^\text{el}(x) = \tr[\b{n}^\text{el}(x)]$ is its associated density. The direct-Breit-type (DB) and exchange-Breit-type (XB) contributions are
\begin{eqnarray}
{\cal E}^{\text{vp},(1),\text{DB}}_\text{b} = -\frac{1}{c^2} \int_{-\infty}^\infty j^\text{el}(x) j^\text{vp}(x) \d x,
\label{Evp1bDB}
\end{eqnarray}
and
\begin{eqnarray}
{\cal E}^{\text{vp},(1),\text{XB}}_\text{b} = \frac{1}{c^2} \int_{-\infty}^\infty \tr[ \b{j}^\text{el}(x) \b{j}^\text{vp}(x) ] \d x,
\label{Evp1bXB}
\end{eqnarray}
where $\b{j}^\text{el}(x) = c \bm{\sigma}_1 \b{n}^\text{el}(x)$ and $\b{j}^\text{vp}(x) = c \bm{\sigma}_1 \b{n}^\text{vp}(x)$ are the electronic bound-state and vacuum-polarization current local density matrices, and $j^\text{el}(x) = \tr[\b{j}^\text{el}(x)]$ and $j^\text{vp}(x) = \tr[\b{j}^\text{vp}(x)]$ are their associated current densities. By time-reversal symmetry, the current densities vanish, i.e. $j^\text{el}(x)=j^\text{vp}(x)=0$ (see, e.g., Ref.~\onlinecite{SalSau-PRA-23}), and thus the DB contribution is zero, i.e. ${\cal E}^{\text{vp},(1),\text{DB}}_\text{b}=0$.

Fig.~\ref{fig:lambshift} reports these QED corrections to the bound-state energy as a function of the inverse speed of light $1/c$, using either the Uehling or total vacuum-polarization local density matrix [Eqs.~(\ref{nvpUxmat}) and (\ref{n1vpx})]. Whereas the total first-order QED correction was negative without the Dirac-delta contribution~\cite{AudTou-JCP-23}, it becomes positive after adding it. Thus, the total effect of the vacuum polarization in the present model is to destabilize the bound-state energy. We note that, for the 3D hydrogen-like atom, the total QED energy correction on the ground-state energy is also positive, i.e. the ground state is also destabilized (see, e.g., Ref.~\onlinecite{EidGroShe-PR-01}).

\section{Finite-dimensional approximation to the 1D hydrogen-like Dirac model}
\label{sec:finitedim}

In this section, we study the 1D hydrogen-like Dirac model using a finite plane-wave basis.

\subsection{Calculations in a finite plane-wave basis}
\label{sec:calculations}
We consider a finite-dimensional approximation of the Hilbert space with an infrared (IR) cutoff parameter $L$ and an ultraviolet (UV) cutoff parameter $\Lambda$,
\begin{eqnarray}
\calh^{L,\Lambda} = \text{span} \left( \left\{ \bm{\chi}_{n}^\text{L}\right\}_{|n| \leq n_\text{max}} \cup \left\{ \bm{\chi}_{n}^\text{S} \right\}_{|n| \leq n_\text{max}} \right),
\end{eqnarray}
which is spanned by a basis of large- and small-component plane-wave functions $\bm{\chi}_{n}^\text{L}, \bm{\chi}_{n}^\text{S}: [-L/2,L/2] \to \mathbb{C}^2$
\begin{eqnarray}
\bm{\chi}_{n}^\text{L}(x) = \frac{1}{\sqrt{L}}
\left(\!\begin{array}{c}
e^{\i k_n x } \\
0
\end{array}\!\right)
\;\; \text{and}\;\;
\bm{\chi}_{n}^\text{S}(x) = \frac{1}{\sqrt{L}} \left(\!\begin{array}{c}
0\\
e^{\i k_n x } 
\end{array}\!\right),
\label{}
\end{eqnarray}
with $k_n = 2\pi n/L$ and $n_\text{max} = \lfloor L \Lambda/(2\pi) \rfloor$. Physically, it corresponds to an electron on the interval $[-L/2,L/2]$ with maximal momentum $|k_{n_\text{max}}| \leq \Lambda$. Ultimately, we will be interested in the limits $L \to \infty$ and $\Lambda \to \infty$.

Using Eq.~(\ref{psiDphi}), the 1D hydrogen-like Dirac eigenvalue equation, $\bm{D}_Z\bm{\psi}_p^Z = \varepsilon_p^Z \bm{\psi}_p^Z$, on this finite-dimensional Hilbert space leads to the following matrix eigenvalue equation
\begin{eqnarray}
\begin{pmatrix} mc^2 \b{I} + \b{V} & c\b{P} \\ c\b{P} & -mc^2 \b{I}+ \b{V}\end{pmatrix}
\begin{pmatrix} \b{c}_p^\text{L} \\ \b{c}_p^\text{S} \end{pmatrix}
= \varepsilon_p^Z
\begin{pmatrix} \b{c}_p^\text{L} \\ \b{c}_p^\text{S} \end{pmatrix},
\label{DPsi=EPsimat}
\end{eqnarray}
where $\b{c}_p^\text{L}$ and $\b{c}_p^\text{S}$ are $(2n_\text{max}+1)$-component vectors, and $\b{I}$, $\b{P}$, and $\b{V}$ are $(2n_\text{max}+1)\times(2n_\text{max}+1)$ matrices with elements
\begin{eqnarray}
\b{I}_{n,m} = \braket{\bm{\chi}_{n}^\text{L}}{\bm{\chi}_{m}^\text{L}}_{\!_L} = \braket{\bm{\chi}_{n}^\text{S}}{\bm{\chi}_{m}^\text{S}}_{\!_L} = \delta_{n,m},
\end{eqnarray}
\begin{eqnarray}
\b{P}_{n,m} = \braket{\bm{\chi}_{n}^\text{L}}{\bm{\sigma}_1 p_x\, \bm{\chi}_{m}^\text{S}}_{\!_L} = \braket{\bm{\chi}_{n}^\text{S}}{\bm{\sigma}_1 p_x\, \bm{\chi}_{m}^\text{L}}_{\!_L} = k_n \delta_{n,m},
\end{eqnarray}
and
\begin{eqnarray}
\b{V}_{n,m} = -Z \bm{\chi}_{n}^{\text{L}\dagger}(0) \bm{\chi}_{m}^\text{L}(0) = -Z\bm{\chi}_{n}^{\text{S}\dagger}(0) \bm{\chi}_{m}^\text{S}(0) = -\frac{Z}{L},
\end{eqnarray}
where $\braket{.\,}{.}_{\!_L}$ denotes here the standard inner product in $L^2([-L/2,L/2],\mathbb{C}^2)$. 
Note that, since $\bm{\sigma}_1 p_x\, \bm{\chi}_{n}^\text{L} = k_n \bm{\chi}_{n}^\text{S}$ and $\bm{\sigma}_1 p_x\, \bm{\chi}_{n}^\text{S} = k_n \bm{\chi}_{n}^\text{L}$, the plane-wave basis satisfies the kinetic-balance condition in all its variants (see, e.g., Refs.~\onlinecite{StaHav-JCP-84,Kut-IJQC-84,ShaTupYerPluSof-PRL-04,SunLiuKut-TCA-11,SalSau-S-20,GraQui-A-22,SalSau-PRA-23}). This ensures that, as $c\to \infty$, the non-relativistic limit with the same basis is correctly reached. Moreover, by analogy with the 3D case, since the plane-wave basis is adapted to the free Dirac operator $\bm{D}_0$, it is expected to avoid any spectral pollution~\cite{LewSer-PLMS-09,LewSer-INC-14}.

After solving Eq.~(\ref{DPsi=EPsimat}), we obtain $4n_\text{max}+2$ eigenfunctions of the form
\begin{eqnarray}
\bm{\psi}_p^Z(x) = \sum_{n=-n_\text{max}}^{n_\text{max}} c_{p,n}^\text{L} \bm{\chi}_n^\text{L}(x) + \sum_{n=-n_\text{max}}^{n_\text{max}} c_{p,n}^\text{S} \bm{\chi}_n^\text{S}(x),
\end{eqnarray}
which can be partitioned into a set of positive-energy states (PS) $\{ \bm{\psi}_p^Z \}_{p \in \text{PS}}$ and a set of negative-energy states (NS) $\{ \bm{\psi}_p^Z \}_{p \in \text{NS}}$. The vacuum-polarization local density matrix can then be calculated as, for given IR cutoff parameter $L$ and UV cutoff parameter $\Lambda$,
\begin{eqnarray}
\b{n}^\text{vp}_{L,\Lambda}(x) =  \sum_{p \in \text{NS}} \bm{\psi}_p^Z(x) \bm{\psi}_p^{Z\dagger}(x) -  \sum_{p \in \text{NS}} \bm{\psi}_p^0(x) \bm{\psi}_p^{0\dagger}(x),
\label{nvpxbasis}
\end{eqnarray}
where $\{\bm{\psi}_p^0\}$ are the eigenfunctions for $Z=0$ calculated in the same basis. The corresponding vacuum-polarization density is $n^\text{vp}_{L,\Lambda}(x) = \tr[\b{n}^\text{vp}_{L,\Lambda}(x)]$. We also calculate in the finite basis the Uehling vacuum-polarization local density matrix $\b{n}^\text{vp,(1)}_{L,\Lambda}(x)$ and the Uehling vacuum-polarization density $n^\text{vp,(1)}_{L,\Lambda}(x)$.

\subsection{Convergence of the bound-state energy and eigenfunction}
\label{sec:convbound}

\begin{figure*}
        \centering
        \includegraphics[width=0.32\textwidth,angle=-90]{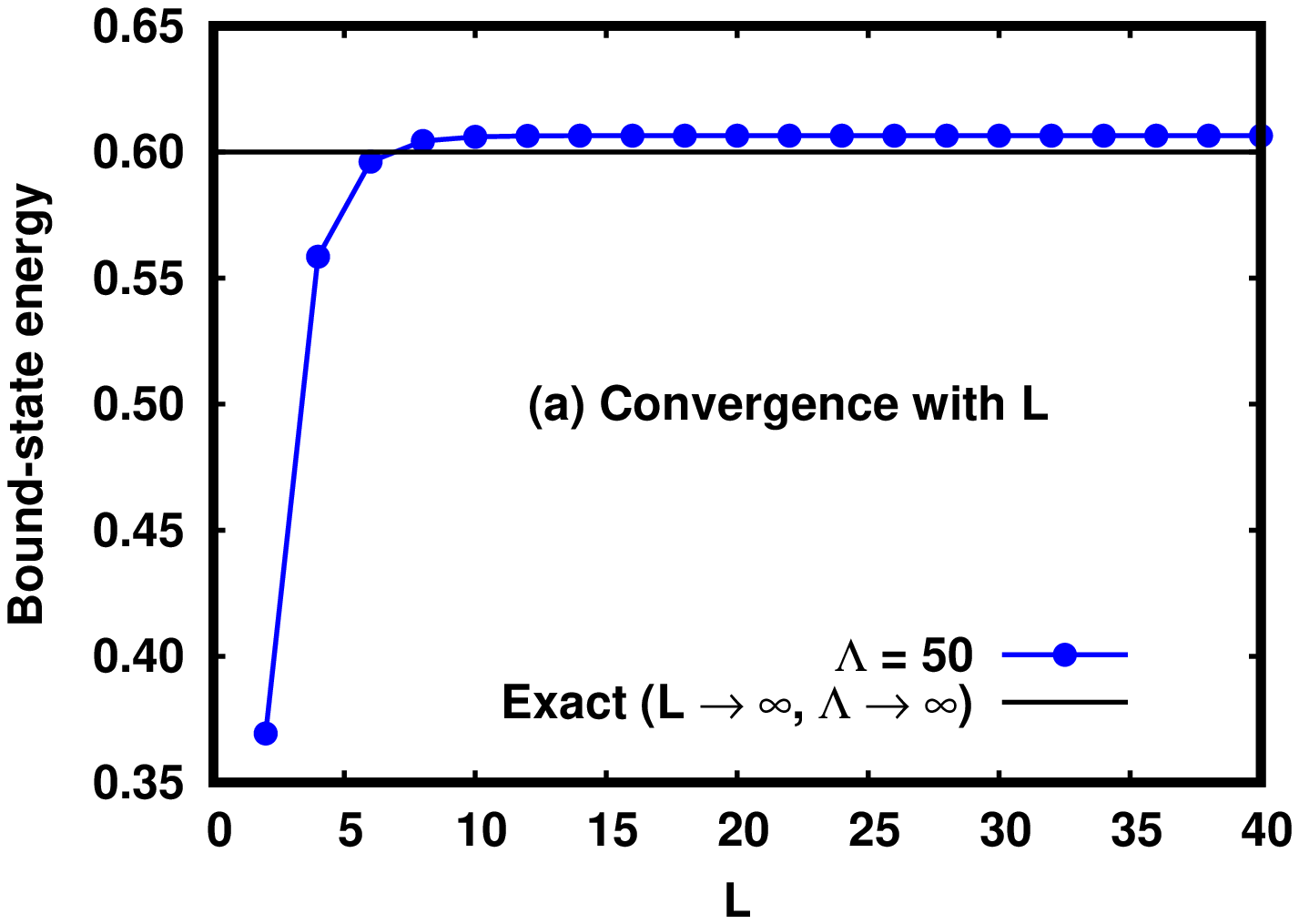}
        \includegraphics[width=0.32\textwidth,angle=-90]{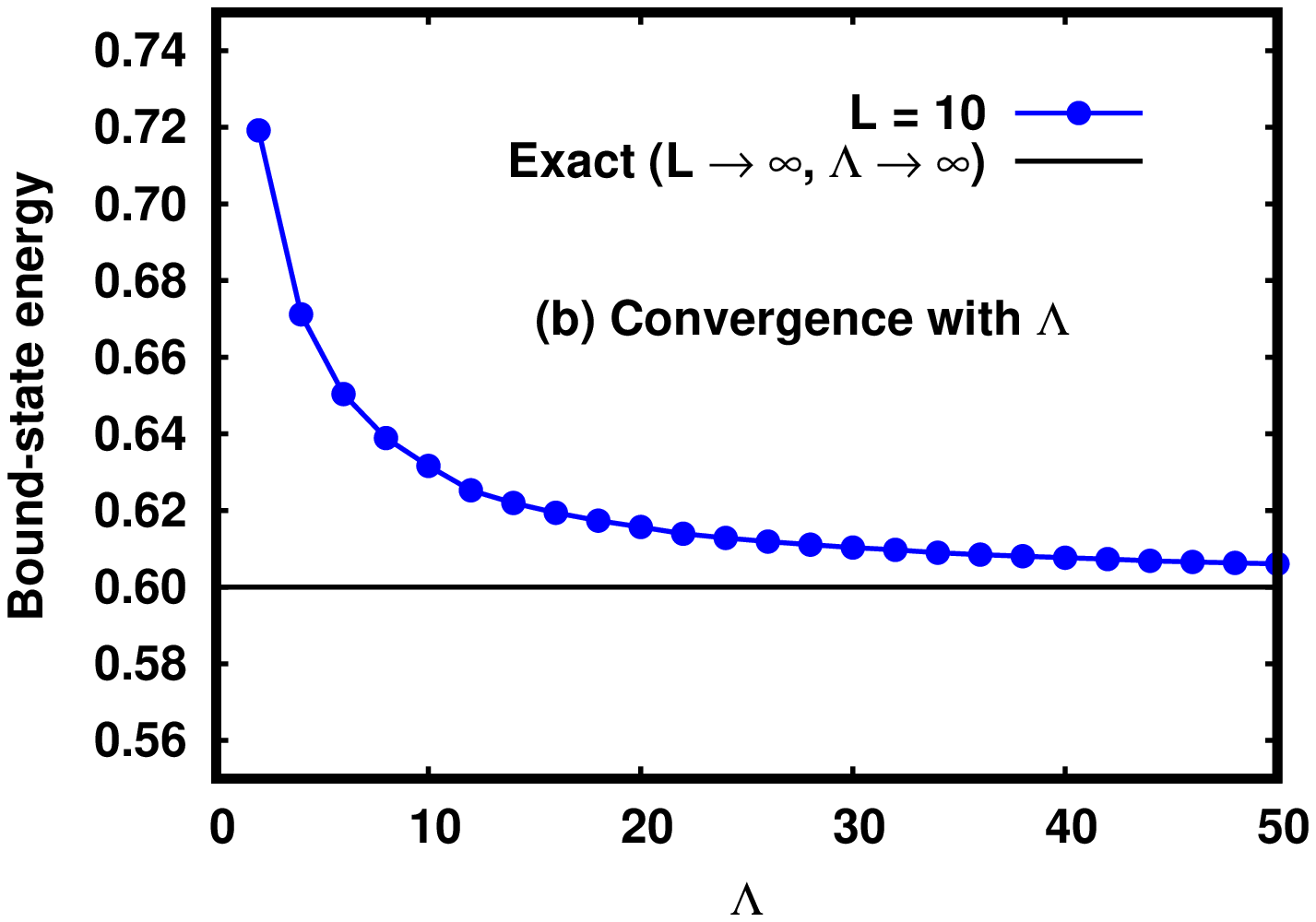}
        \caption{Convergence of the bound-state energy of the 1D hydrogen-like Dirac model with a plane-wave basis as a function of (a) the IR cutoff parameter $L$ and (b) the UV cutoff parameter $\Lambda$ for $m=c=Z=1$. The exact value in the limits $L\to \infty$ and $\Lambda \to \infty$ is $\varepsilon_\text{b} = 0.6$.}
        \label{fig:boundstateenergy}
\end{figure*}

Figure~\ref{fig:boundstateenergy} reports the convergence of the bound-state energy as a function of the IR cutoff parameter $L$ and the UV cutoff parameter $\Lambda$. As $L\to\infty$ and $\Lambda \to \infty$, the bound-state energy calculated in the plane-wave basis correctly converges to the exact value in Eq.~(\ref{epsilon0tilde}). This supports the fact that the basis calculations based on Eq.~(\ref{psiDphi}) corresponds to the self-adjoint realization of the Hamiltonian $\bm{D}_Z$ that we have selected via Eqs.~(\ref{Dpsi=D0psi})-(\ref{1stboundarycond}). Numerically, we find that the bound-state energy converges exponentially as $L\to\infty$, and roughly as $1/\Lambda$ as $\Lambda \to \infty$. This is expected based on the theoretical analysis in Appendix~\ref{app:rate}. 

\begin{figure*}
        \centering
        \includegraphics[width=0.32\textwidth,angle=-90]{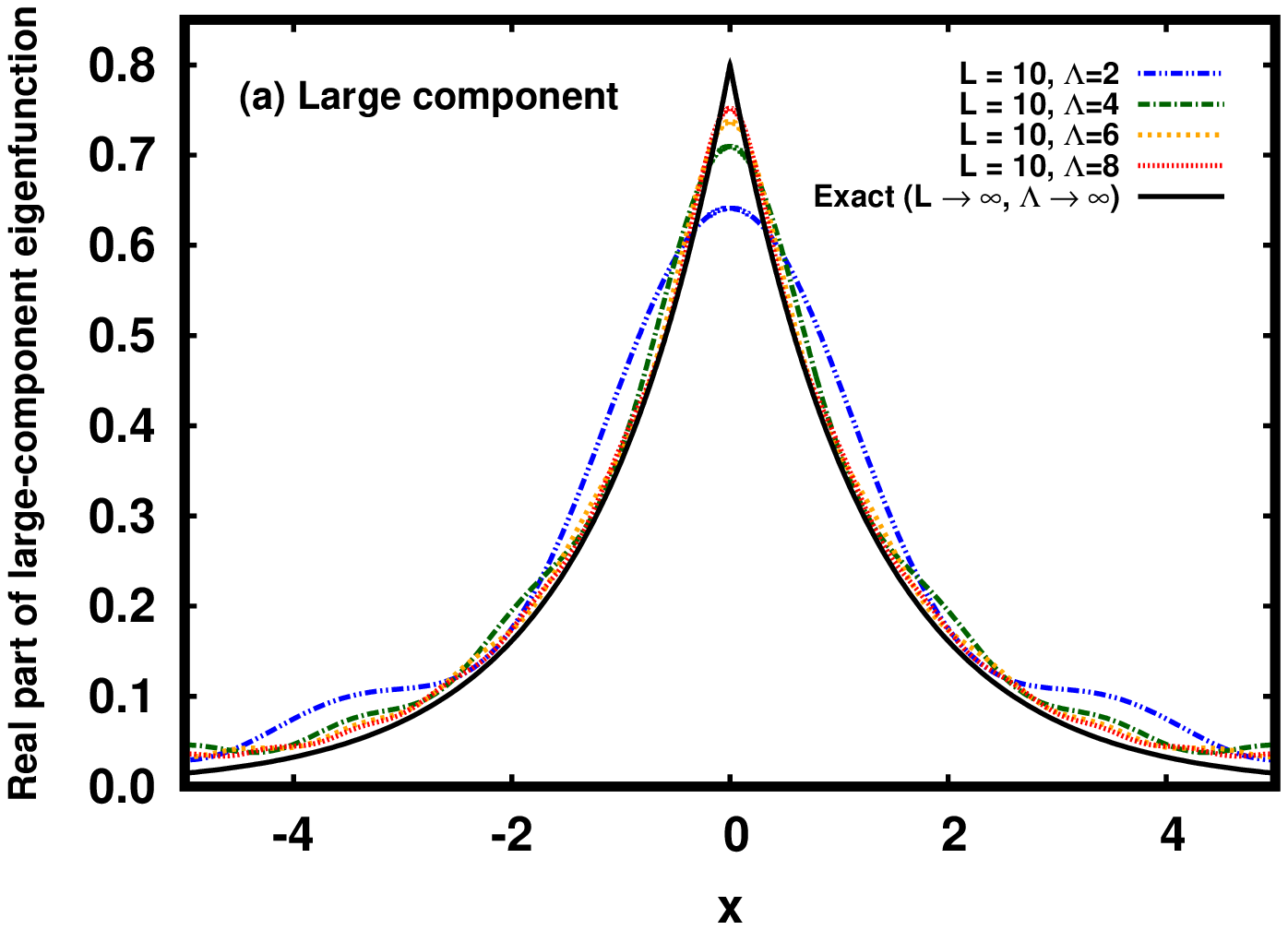}
        \includegraphics[width=0.32\textwidth,angle=-90]{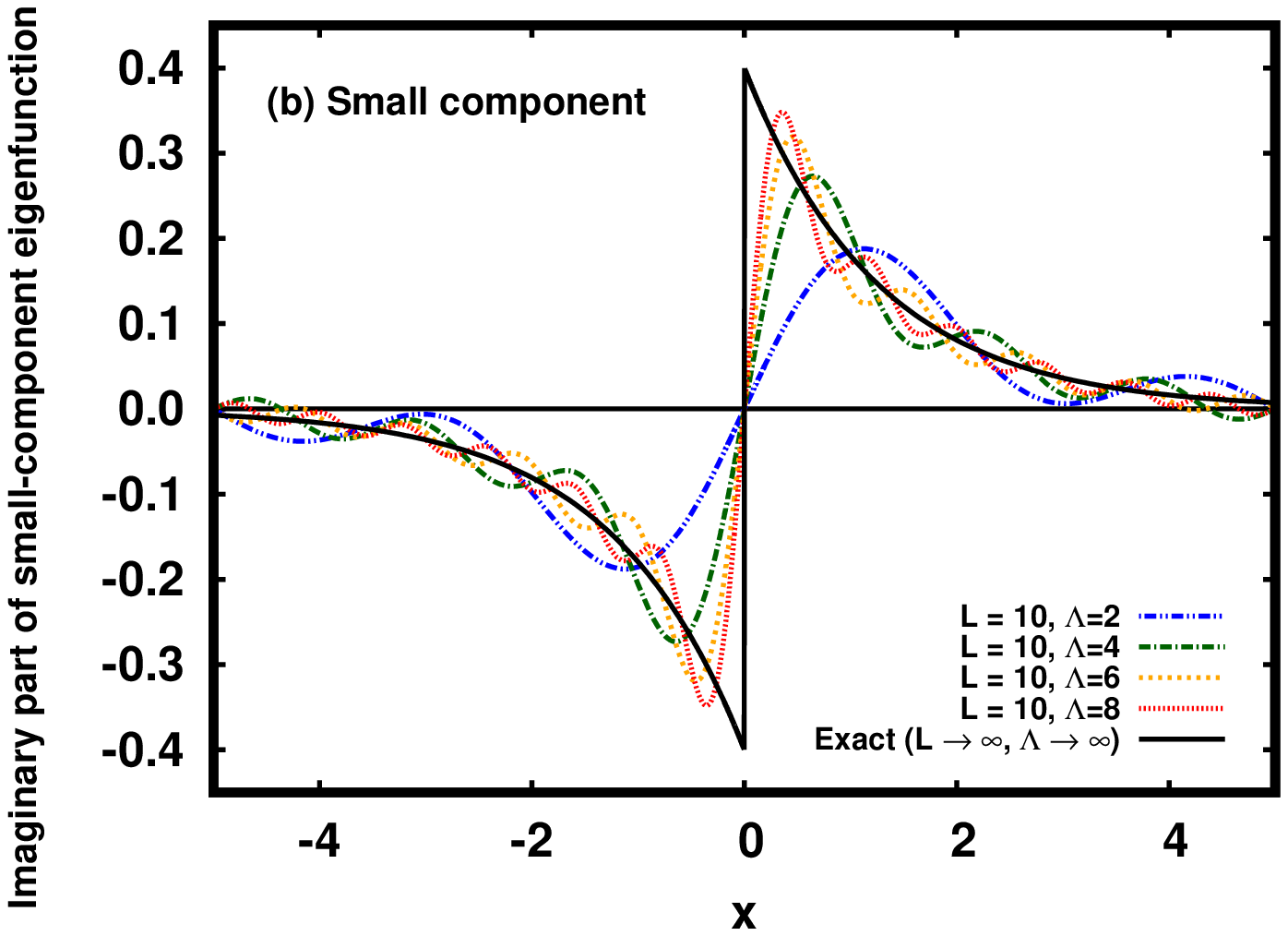}
        \caption{Convergence of the (a) large component and (b) small component of the bound-state eigenfunction of the 1D hydrogen-like Dirac model with a plane-wave basis as a function of the UV cutoff parameter $\Lambda$ for an IR cutoff parameter $L=10$ and $m=c=Z=1$. The exact eigenfunction [Eq.~(\ref{psi0tilde})] corresponds to the limits $L\to \infty$ and $\Lambda \to \infty$ .}
        \label{fig:boundstatewavefunction}
\end{figure*}

Figure~\ref{fig:boundstatewavefunction} reports the convergence of the large and small components of the bound-state eigenfunction as a function of the IR cutoff parameter $L$ and the UV cutoff parameter $\Lambda$. The large component of the exact eigenfunction [see Eq.~(\ref{psi0tilde})] has a derivative discontinuity at $x=0$, and consequently the convergence with $\Lambda$ is slow near $x=0$. The small component of the exact eigenfunction [see Eq.~(\ref{psi0tilde})] has a discontinuity at $x=0$, in addition of having the same derivative discontinuity at $x=0$ as the large component, and consequently the convergence with respect to $\Lambda$ near $x=0$ is even slower than for the large component. Note that this discontinuity at $x=0$ of the exact small-component eigenfunction cannot be reproduced in the finite plane-wave basis, since the expansion of an odd function in a basis of continuous functions necessarily gives zero at $x=0$. This implies that the finite plane-wave basis always incorrectly gives a vanishing small-component contribution to the bound-state density at $x=0$. As seen from the analysis in Appendix~\ref{app:rate}, the convergence of the small-component eigenfunction is the limiting factor in the convergence of the bound-state energy as $\Lambda \to \infty$. Naively, this last fact might lead one to think that a faster convergence of the bound-state energy with respect to $\Lambda$ would be obtained in the non-relativistic limit since only the large component survives in this limit. In fact, it can be shown that, in the non-relativistic limit, the bound-state energy still converges as $1/\Lambda$ as $\Lambda \to \infty$. This is because the non-relativistic Hamiltonian involves now the second-order derivative of the large-component eigenfunction (see, also, Ref.~\onlinecite{TraGinTou-JCP-22} for a related discussion on the basis convergence of the non-relativistic bound-state energy).

\subsection{Convergence of the vacuum-polarization density}

\begin{figure}
        \centering
        \includegraphics[width=0.32\textwidth,angle=-90]{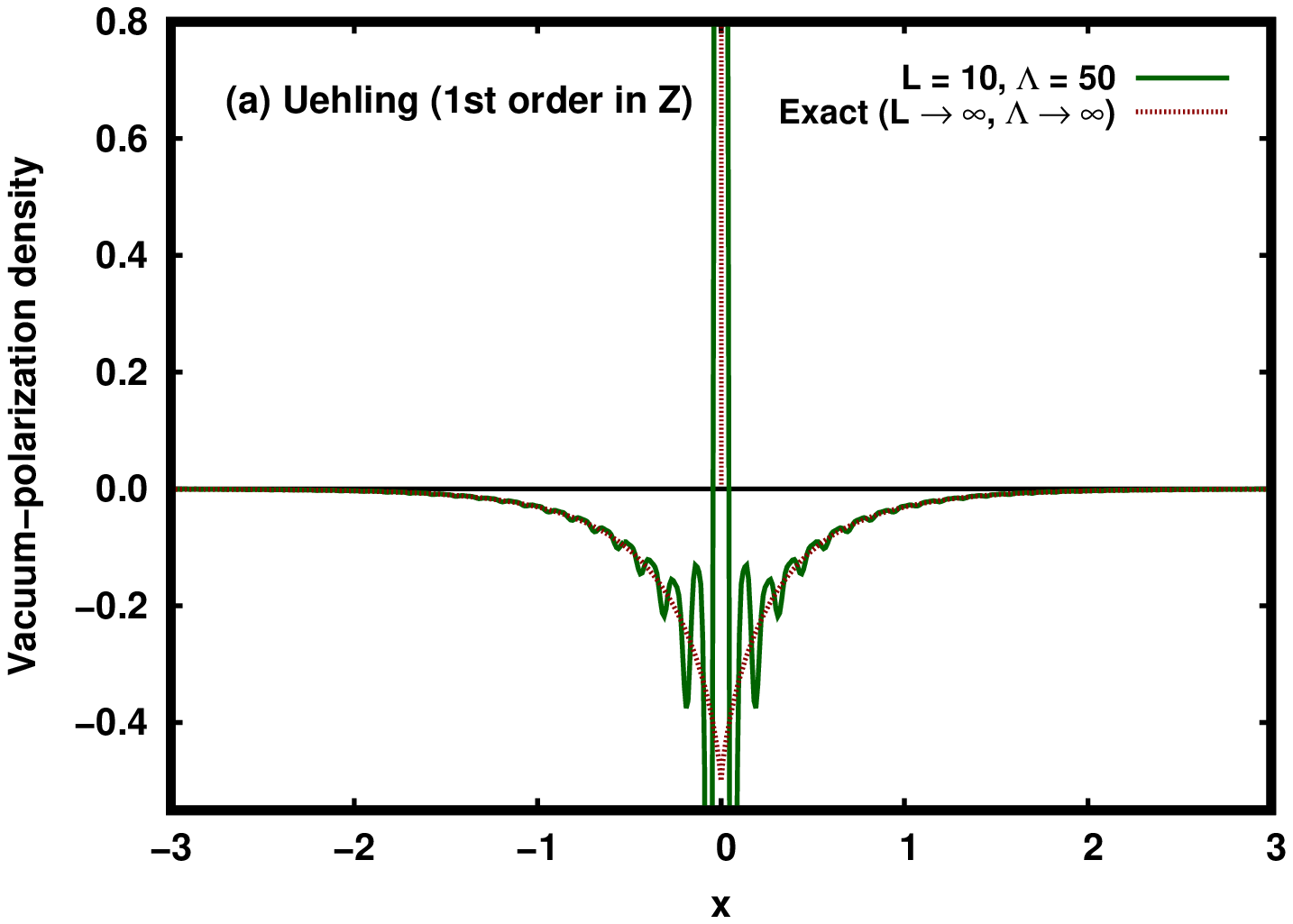}
        \includegraphics[width=0.32\textwidth,angle=-90]{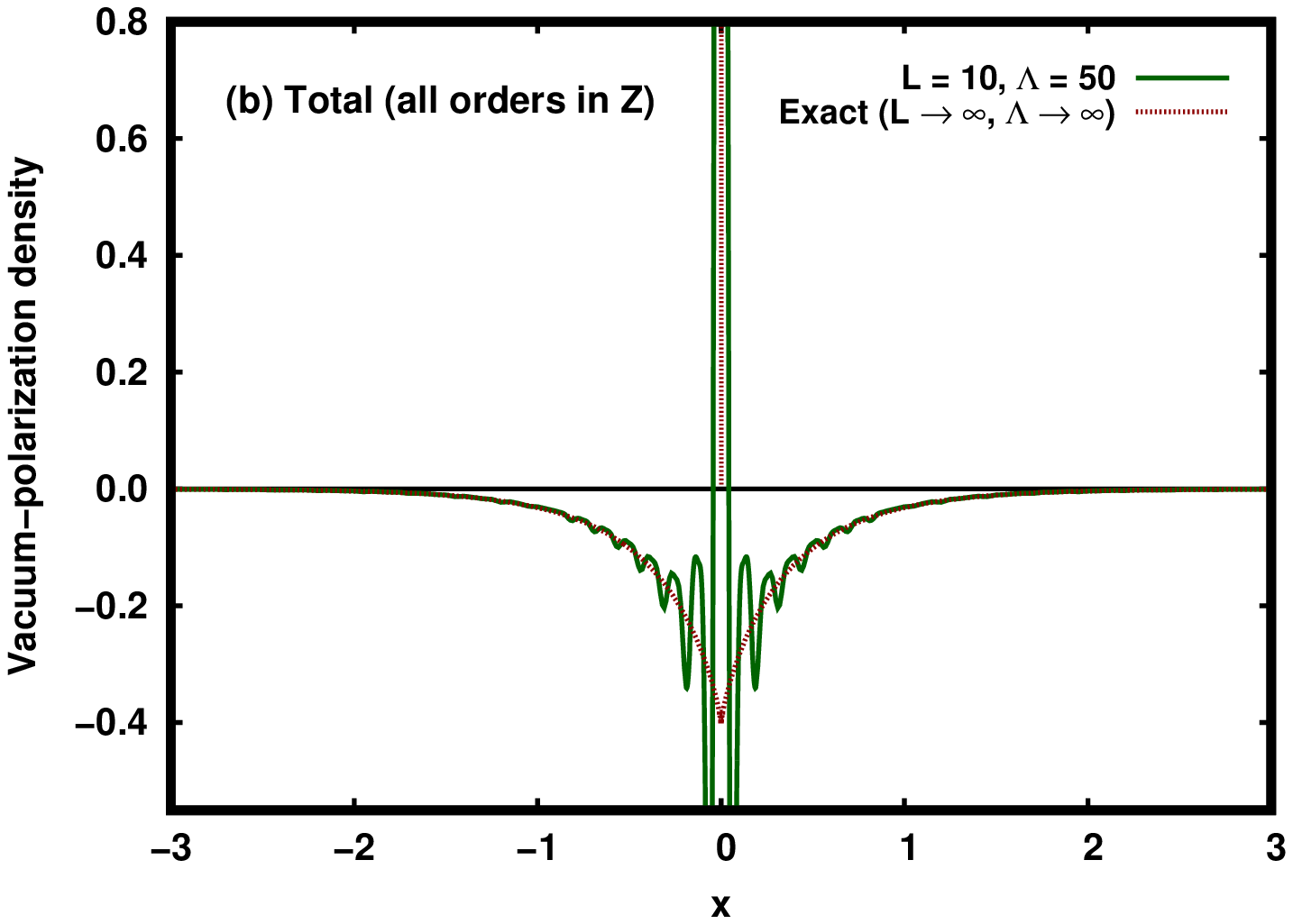}
        \caption{(a) Uehling and (b) total vacuum-polarization densities for $m = c = Z = 1$, calculated exactly [Eqs.~(\ref{nvp1x}) and~(\ref{nvpx})] and with a plane-wave basis for a IR cutoff parameter $L=10$ and a UV cutoff parameter $\Lambda = 50$ [from Eq.~(\ref{nvpxbasis})].}
        \label{fig:VPDpw}
\end{figure}

We now discuss the convergence of the vacuum-polarization density $n^{\vp}_{L,\Lambda}(x)$ calculated from Eq.~(\ref{nvpxbasis}) with a finite plane-wave basis. The Uehling and total vacuum-polarization densities calculated with a plane-wave basis with a IR cutoff parameter $L=10$ and a UV cutoff parameter $\Lambda = 50$ are reported in Fig.~\ref{fig:VPDpw}, and compared with the exact ones [Eqs.~(\ref{nvp1x}) and~(\ref{nvpx})]. The plane-wave basis calculation reproduces well the vacuum-polarization density for large enough $x$. However, the plane-wave basis generates large oscillations near $x=0$ while trying to reproduce the Dirac-delta contribution, resulting in an effectively impossible pointwise convergence of the vacuum-polarization density near $x=0$.

\begin{figure}
        \centering
        \includegraphics[width=0.32\textwidth,angle=-90]{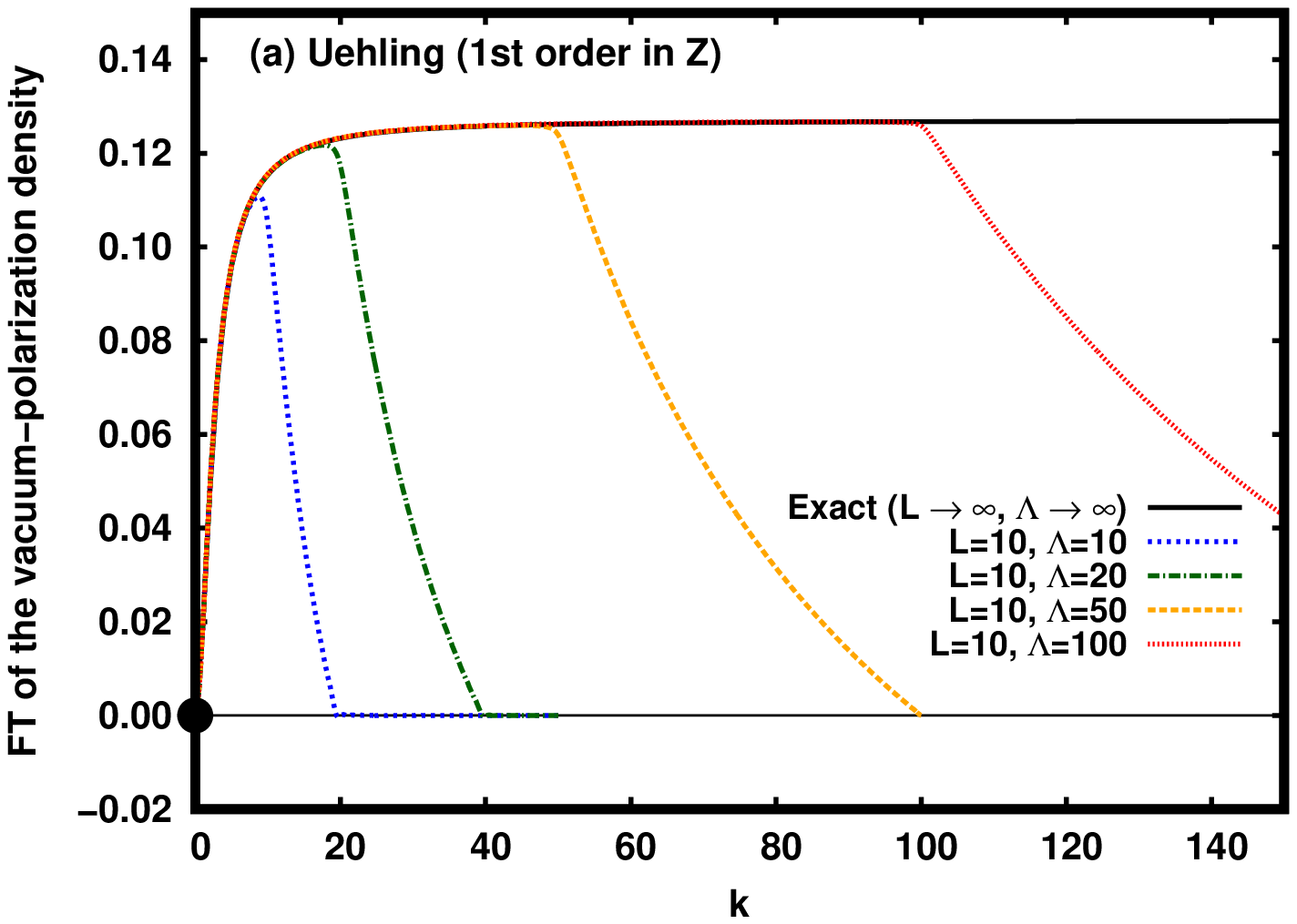}
        \includegraphics[width=0.32\textwidth,angle=-90]{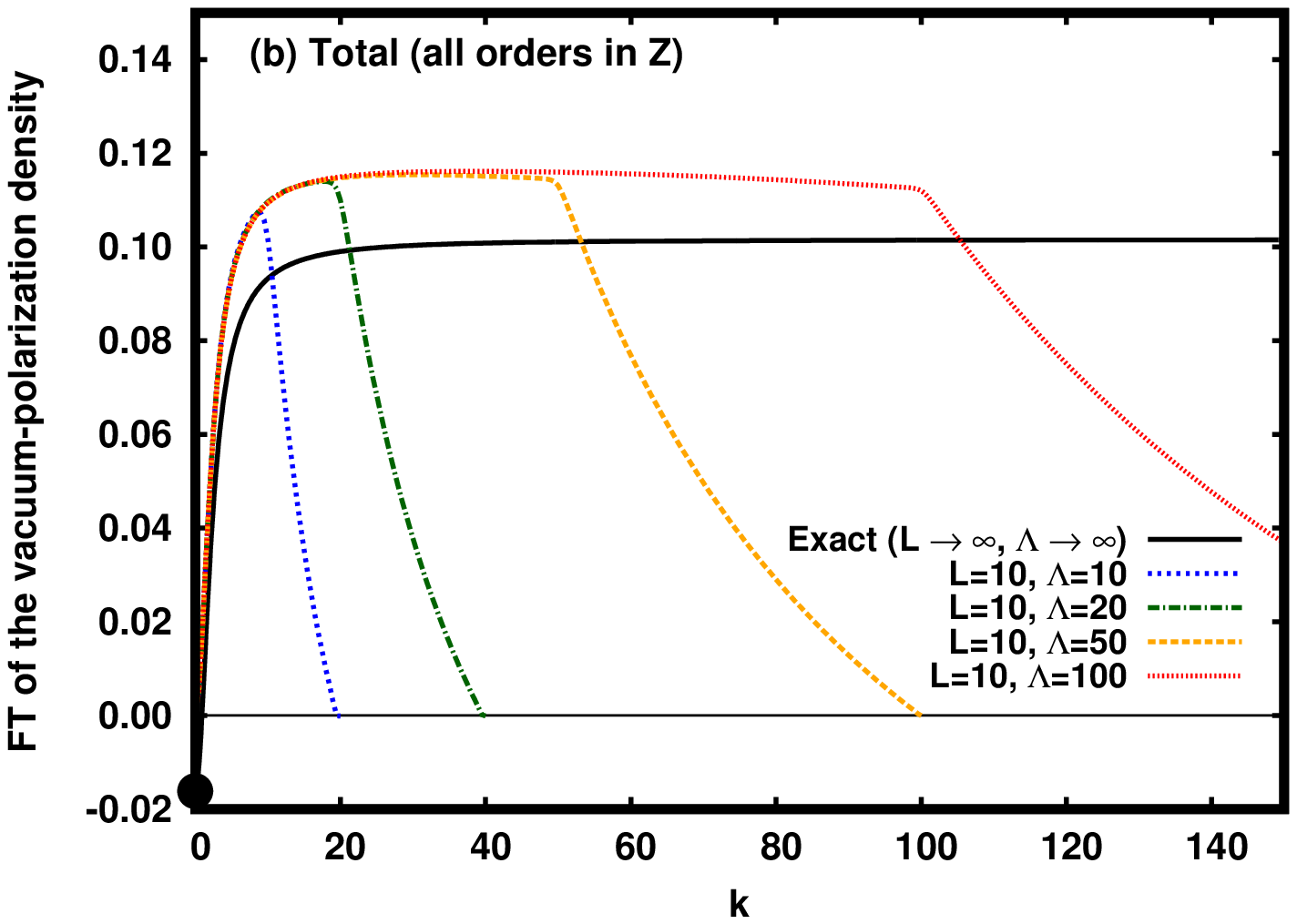}
        \caption{Fourier transform of the (a) Uehling and (b) total vacuum-polarization densities calculated exactly [Eqs.~(\ref{nvp1x}) and~(\ref{nvpx})] and with a plane-wave basis for a IR cutoff parameter $L=10$ and different UV cutoff parameters $\Lambda$ for $m = c = Z = 1$. The black dot indicates the beginning of the exact curve (for $L\to \infty$ and $\Lambda \to \infty$).}
        \label{fig:FT}
\end{figure}

To analyze this convergence problem, we report in Fig.~\ref{fig:FT} the Fourier transform of the vacuum-polarization density calculated with the plane-wave basis
\begin{eqnarray}
\hat{n}^\text{vp}_{L,\Lambda}(k) =  \frac{1}{\sqrt{2\pi}} \int_{-L/2}^{L/2} n^{\vp}_{L,\Lambda}(x) e^{-\i k x} \d x.
\label{nvpLGk}
\end{eqnarray}
To reproduce the Dirac-delta contribution to the vacuum-polarization density in position space, $\hat{n}^\text{vp}_{L,\Lambda}(k)$ should tend to a constant as $k\to\infty$. However, with a finite UV cutoff parameter $\Lambda$, the Fourier transform $\hat{n}^\text{vp}_{L,\Lambda}(k)$ is only able to approximately converge to a constant for $k \lesssim \Lambda$, but for $k \gtrsim \Lambda$ quickly decays to reach zero at $k=2\Lambda$. Note that the reason why the Fourier transform of the vacuum-polarization density is not zero for $\Lambda \leq k \leq 2\Lambda$ is that, according to Eq.~(\ref{nvpxbasis}), the vacuum-polarization density involves the product of two plane waves, each with momenta going up to $\Lambda$. However, $\hat{n}^\text{vp}_{L,\Lambda}(k)$ cannot be expected to be accurate for the momentum range $\Lambda \leq k \leq 2\Lambda$ since contributions from plane waves with momenta greater than $\Lambda$ are missing.

For the Uehling vacuum-polarization density, the basis calculation produces a curve of $\hat{n}^\text{vp,(1)}_{L,\Lambda}(k)$ which is nearly superimposed with the curve of the exact $\hat{n}^\text{vp,(1)}(k)$ for $k \lesssim \Lambda$. However, for the total vacuum-polarization density, the basis calculation produces a curve of $\hat{n}^\text{vp}_{L,\Lambda}(k)$ which is shifted upward with respect to the curve of the exact $\hat{n}^\text{vp}(k)$. This is due to the fact that the exact total vacuum-polarization density does not integrate to zero, as explained in Section~\ref{sec:totalvpd}. Hence, the Fourier transform of the exact total vacuum-polarization density is not zero at $k=0$, i.e. $\hat{n}^\text{vp}(0)={\cal N}^\text{vp}/\sqrt{2\pi} \not = 0$. This behavior cannot be reproduced with our finite plane-wave basis calculation. Indeed, since in our finite plane-wave basis, we obtain the same number of negative-energy states for a non-vanishing nuclear charge $Z< 2c$ and $Z=0$, the total vacuum-polarization densities necessarily integrates to zero for any finite cutoff parameters $L$ and $\Lambda$, i.e. $\int_{-\infty}^\infty n^{\vp}_{L,\Lambda} (x) \d x = 0$ and $\hat{n}^\text{vp}_{L,\Lambda}(0)=0$
.

\begin{figure*}
        \centering
        \includegraphics[width=0.32\textwidth,angle=-90]{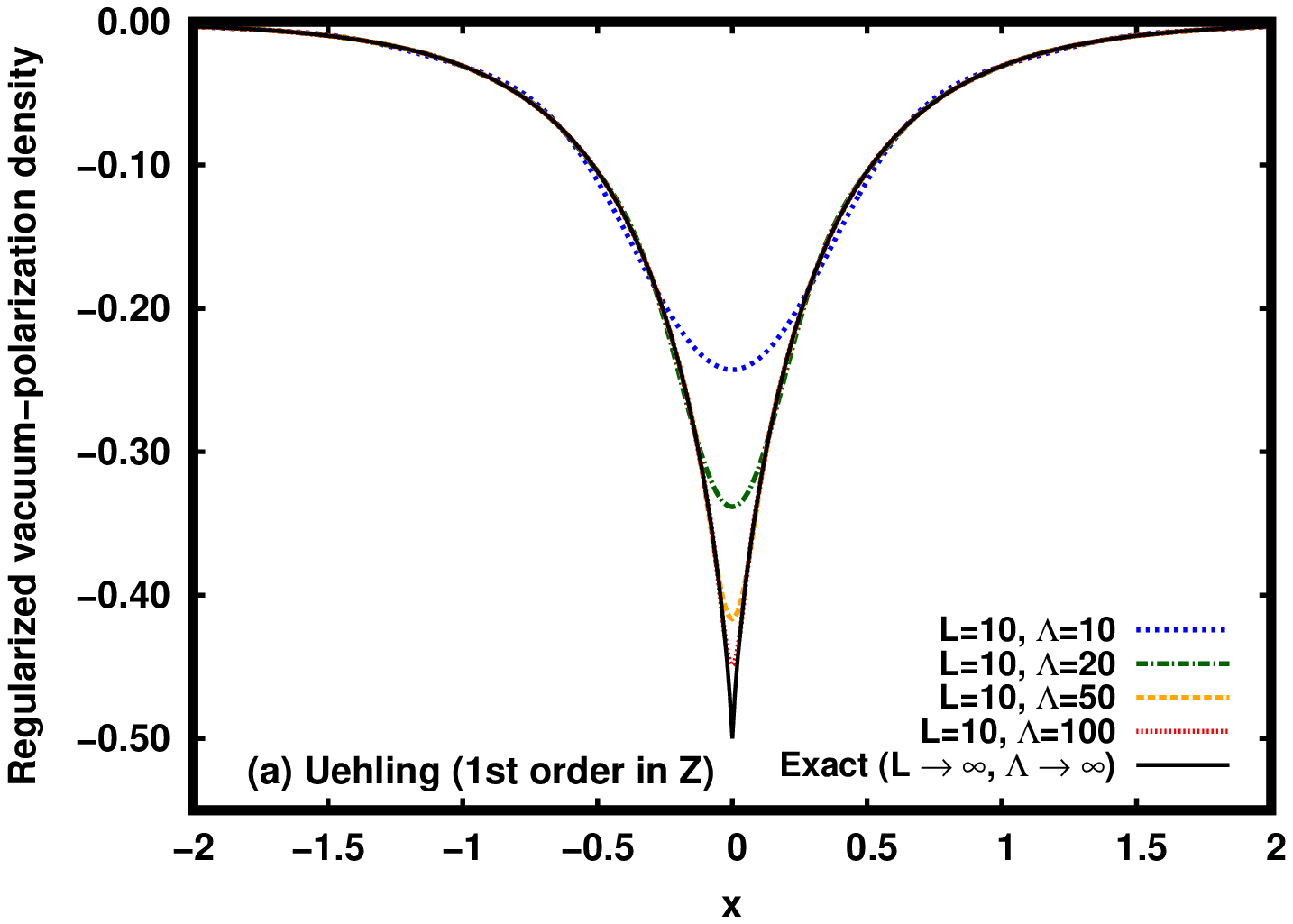}
        \includegraphics[width=0.32\textwidth,angle=-90]{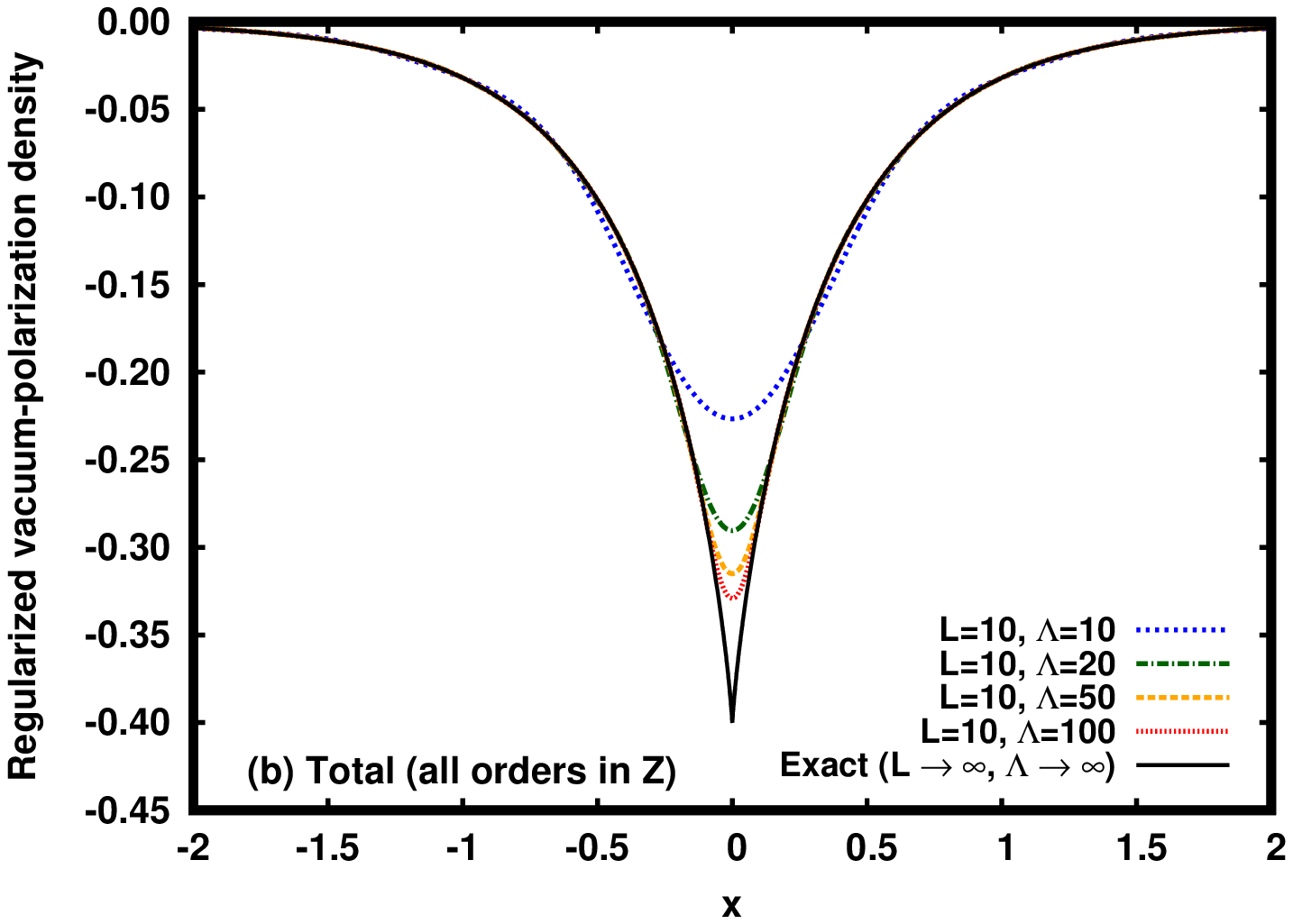}
        \caption{Regularized (a) Uehling and (b) total vacuum-polarization densities [Eq.~(\ref{nvpLGxreg})] for $m = c = Z = 1$ calculated in a plane-wave basis for a IR cutoff parameter $L=10$ and different UV cutoff parameters $\Lambda$, compared to the exact regular part of the Uehling and total vacuum-polarization densities.}
        \label{fig:nvpLGxreg}
\end{figure*}

The above analysis suggests the following simple scheme to extract the regular part of the Uehling or total vacuum-polarization density in a finite basis. For given $L$ and $\Lambda$, we select a maximal momentum $k_\text{max}$, and truncate $\hat{n}^\text{vp}_{L,\Lambda}(k) - \hat{n}^\text{vp}_{L,\Lambda}(k_\text{max})$ for $|k|> k_\text{max}$, i.e.
\begin{eqnarray}
\hat{n}^\text{vp}_{\text{reg},L,\Lambda}(k) = \left[ \hat{n}^\text{vp}_{L,\Lambda}(k) - \hat{n}^\text{vp}_{L,\Lambda}(k_\text{max}) \right] \theta(k_\text{max} -|k|).
\label{nvpLGkreg}
\end{eqnarray}
By construction, $\hat{n}^\text{vp}_{\text{reg},L,\Lambda}(k)$ is thus zero for $|k| > k_\text{max}$. In practice, we choose for $k_\text{max}$ the value of $k$ at which $\hat{n}^\text{vp}_{L,\Lambda}(k)$ reaches its maximum, i.e.
\begin{eqnarray}
k_\text{max} = \argmax{k\in[0,2\Lambda]} \; \hat{n}^\text{vp}_{L,\Lambda}(k).
\label{kmax}
\end{eqnarray}
The values of $k_\text{max}$ obtained in this way are close to $\Lambda$. The choice of $k_\text{max}$ in Eq.~(\ref{kmax}), instead of the simpler choice $k_\text{max} = \Lambda$, has the advantage that it leads to $\hat{n}^\text{vp}_{\text{reg},L,\Lambda}$ having a continuous vanishing derivative at $k=k_\text{max}$. We finally take the inverse Fourier transform  of $\hat{n}^\text{vp}_{\text{reg},L,\Lambda}(k)$, i.e.
\begin{eqnarray}
n^\text{vp}_{\text{reg},L,\Lambda}(x) =  \frac{1}{\sqrt{2\pi}} \int_{-k_\text{max}}^{k_\text{max}} \hat{n}^{\vp}_{\text{reg},L,\Lambda}(k) e^{\i k x} \d k,
\label{nvpLGxreg}
\end{eqnarray}
resulting in a regularized vacuum-polarization density where the approximation of the Dirac-delta contribution in the finite basis has been removed. This quantity should then converge to the regular part of the total vacuum-polarization density as $L\to\infty$ and $\Lambda\to\infty$, i.e. $\lim_{L\to\infty} \lim_{\Lambda\to\infty} n^\text{vp}_{\text{reg},L,\Lambda}(x) = n^{\text{vp}}_\text{reg}(x)$. 

The regularized vacuum-polarization density $n^\text{vp}_{\text{reg},L,\Lambda}(x)$ obtained by the above scheme is reported in Fig.~\ref{fig:nvpLGxreg} for different UV cutoff parameters $\Lambda$. We see that our regularization scheme has properly eliminated the large oscillations near $x=0$ and the regularized vacuum-polarization density converges indeed toward the regular part of the exact vacuum-polarization density as the UV cutoff is increased. However, this convergence remains slow near $x=0$ due to the fact that the regular part of the exact vacuum-polarization density has an infinite derivative discontinuity at $x=0$.

Once the regular part of the vacuum-polarization density in the basis has been regularized, one can add the Dirac-delta contribution to obtain the final approximation of the vacuum-polarization density from the finite basis calculation
\begin{eqnarray}
\tilde{n}^{\text{vp}}_{L,\Lambda}(x) = {\cal N}_{\text{0}}^\text{vp}\, \delta(x) + n^\text{vp}_{\text{reg},L,\Lambda}(x).
\label{ntvpxLG}
\end{eqnarray}
In the case of the Uehling vacuum-polarization density, we can also extract the coefficient of the Dirac-delta contribution from the basis calculation, ${\cal N}_{\text{0}}^\text{vp,(1)} \approx {\cal N}_{\text{0},L,\Lambda}^\text{vp,(1)}$ where ${\cal N}_{\text{0},L,\Lambda}^\text{vp,(1)}$ is defined as the opposite of the integral of the regular part
\begin{eqnarray}
{\cal N}^\text{vp,(1)}_{0,L,\Lambda} &=&  -\int_{-L/2}^{L/2} n^\text{vp,(1)}_{\text{reg},L,\Lambda}(x) \d x 
\nonumber\\
&=& -\sqrt{2\pi}  \; \hat{n}^\text{vp,(1)}_{\text{reg},L,\Lambda}(0)
\nonumber\\
&=& \sqrt{2\pi}  \; \hat{n}^\text{vp,(1)}_{L,\Lambda}(k_\text{max}).
\label{Nvp1LGxreg}
\end{eqnarray}
In the case of the total vacuum-polarization density, as far as we can see, there is no way to extract the correct coefficient of the Dirac-delta contribution corresponding to a non-zero integral of the total vacuum-polarization density, since the finite basis calculations always gives a total vacuum-polarization density that integrates to zero. So, in this case, we have to rely on the coefficient of the Dirac-delta contribution obtained from the exact calculation.

\subsection{Convergence of the QED correction to the bound-state energy}
\label{sec:convQEDcorr}

\begin{figure}
        \centering
        \includegraphics[width=0.32\textwidth,angle=-90]{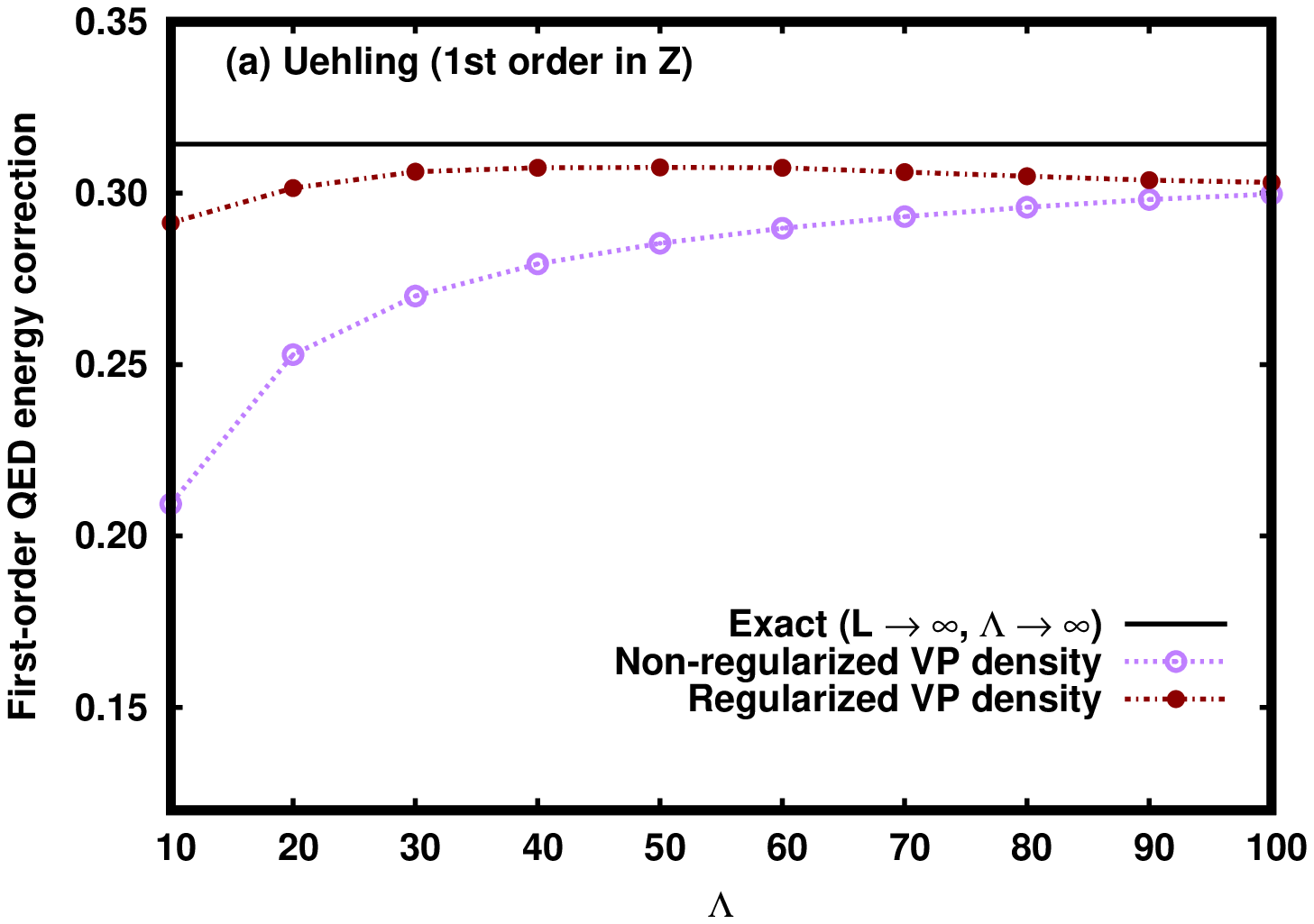}
        \includegraphics[width=0.32\textwidth,angle=-90]{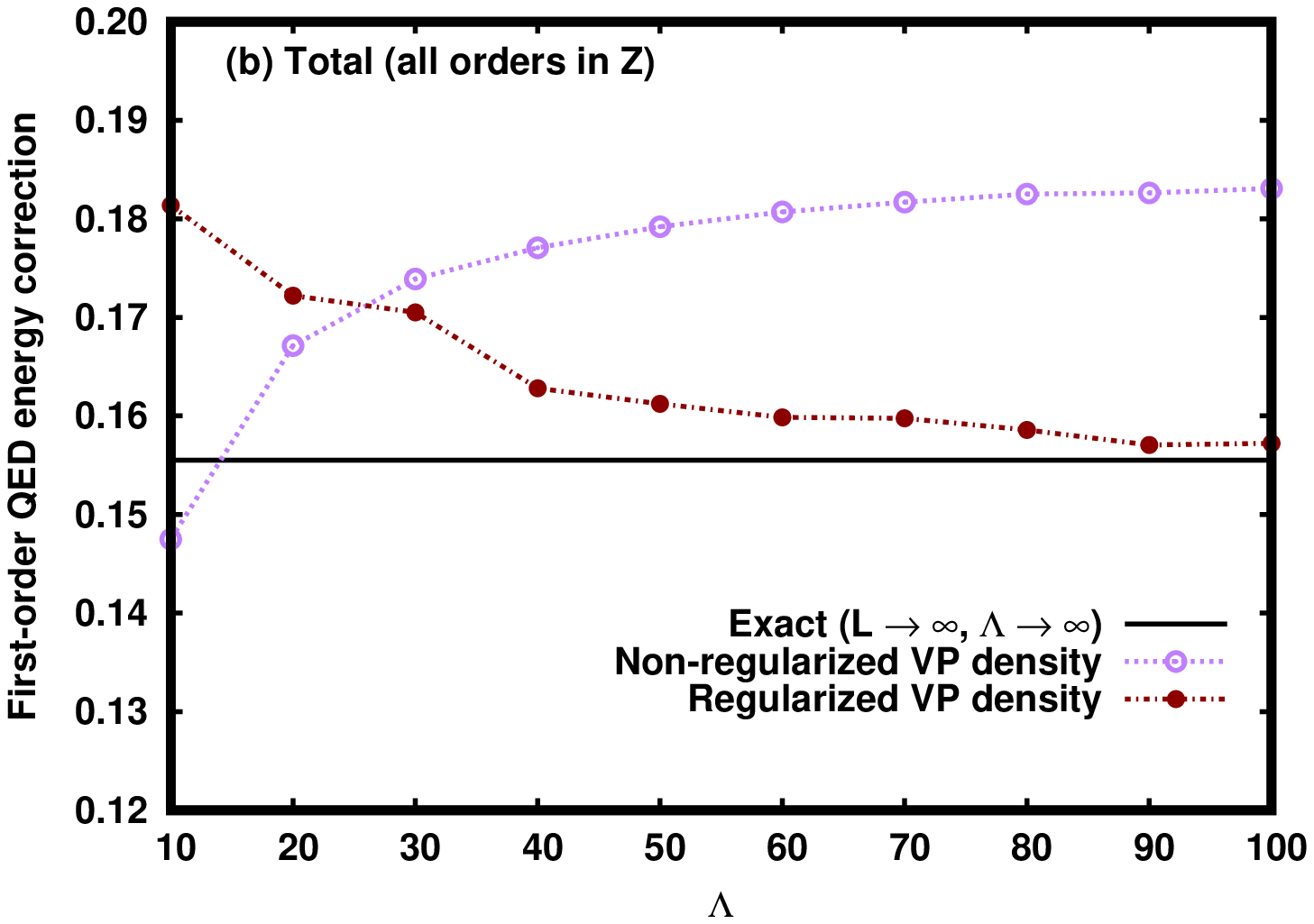}
        \caption{Convergence of the first-order QED correction to the bound-state energy calculated in a plane-wave basis as a function of the UV cutoff parameter $\Lambda$ for a IR cutoff parameter $L=10$ and for $m = c = Z = 1$,  using either the non-regularized vacuum-polarization local density matrix $\b{n}^{\text{vp}}_{L,\Lambda}(x)$ [Eq.~(\ref{nvpxbasis})] or the regularized vacuum-polarization local density matrix $\tilde{\b{n}}^{\text{vp}}_{L,\Lambda}(x)$ [Eq.~(\ref{bnvprenLGx})], for both the (a) Uehling and (b) total vacuum-polarization cases, and the exact bound-state local density matrix $\b{n}^{\text{el}}(x)$.}
        \label{fig:lambshiftpw}
\end{figure}

In Fig.~\ref{fig:lambshiftpw}, we report the convergence of the total first-order QED correction to the bound-state energy in Eq.~(\ref{Evp1b}) calculated in different ways in the plane-wave basis as a function of the UV cutoff parameter $\Lambda$, for both the Uehling and total vacuum-polarization cases. Since the electronic bound-state local density matrix $\b{n}^{\text{el}}_{L,\Lambda}(x)$ calculated in the plane-wave basis has a slow convergence with respect to $\Lambda$ and since we want to focus here on the effect of the vacuum-polarization local density matrix, we use in these calculations the exact electronic bound-state local density matrix $\b{n}^{\text{el}}(x)$. For the vacuum-polarization local density matrix, we use either the non-regularized one $\b{n}^{\text{vp}}_{L,\Lambda}(x)$ in Eq.~(\ref{nvpxbasis}) or the following regularized one
\begin{eqnarray}
\tilde{\b{n}}^{\text{vp}}_{L,\Lambda}(x) = \frac{{\cal N}_{0}^\text{vp}}{2}\, \delta(x) \b{I}_2 + \b{n}^\text{vp}_{\text{reg},L,\Lambda}(x),
\label{bnvprenLGx}
\end{eqnarray}
where $\b{n}^\text{vp}_{\text{reg},L,\Lambda}(x)$ is a regularized vacuum-polarization local density matrix that we define in momentum space as, similarly to Eq.~(\ref{nvpLGkreg}), 
\begin{eqnarray}
\hat{\b{n}}^\text{vp}_{\text{reg},L,\Lambda}(k) = \left[ \hat{\b{n}}^\text{vp}_{L,\Lambda}(k) - \hat{\b{n}}^\text{vp}_{L,\Lambda}(k_\text{max}) \right] \theta(k_\text{max} -|k|),
\label{}
\end{eqnarray}
using the same value for $k_\text{max}$ as the one used for the vacuum-polarization density [see Eq.~(\ref{kmax})].

For the Uehling case, the first-order QED correction obtained from the non-regularized vacuum-polarization local density matrix appears to slowly converge to the exact value, while the first-order QED correction obtained from the regularized vacuum-polarization local density matrix has much smaller errors for small values of $\Lambda$. For the total vacuum-polarization case, the first-order QED correction obtained from the non-regularized vacuum-polarization local density matrix converges to a value larger than the exact value. This is again a manifestation  of the fact that the non-regularized total vacuum-polarization density in the finite basis incorrectly integrates to zero. By contrast, the first-order QED correction obtained from the regularized vacuum-polarization local density matrix, which has the correct coefficient in front of the Dirac-delta contribution, does converge toward the exact value.

\section{Conclusion}
\label{sec:concl}

In this work, we have reexamined the 1D effective QED model of the hydrogen-like atom with delta-potential interactions. We have provided some mathematical details on the definition of this model. We have calculated the exact vacuum-polarization density and local density matrix, using momentum-space Green-function techniques, showing that a Dirac-delta contribution was missed in previous calculations. We have calculated the resulting Lamb-type shift of the bound-state energy. We have also studied the approximation of this model in a finite plane-wave basis, in particular the basis convergence of the bound-state energy and eigenfunction, of the vacuum-polarization density, of the  Lamb-type shift of the bound-state energy. We have shown that it is difficult to converge the vacuum-polarization density with respect to the UV momentum cutoff of the plane-wave basis due to the presence of the Dirac-delta contribution. We have proposed a way of filtering out in momentum space this Dirac-delta contribution in the plane-wave calculations in order to converge the regular part of the vacuum-polarization density in the basis. A possible extension of this work would be to determine the effect of the two-particle interaction~\cite{AudTou-JCP-23} on the vacuum-polarization density and density matrix in the present model and its computational consequences.

We believe that the present work may give some hints on how to perform finite-basis calculations of the vacuum-polarization density of atoms and molecules in the 3D effective QED theory with Coulomb interactions. Indeed, for atoms and molecules, the calculation of the vacuum-polarization density in a finite basis presents similar problems as in the present 1D model. For instance, the vacuum-polarization density of an atom with a point-charge nucleus also contains a Dirac-delta contribution at the nucleus (see, e.g., Refs.~\onlinecite{HaiSie-CMP-03,GreRei-BOOK-09}). A similar regularization technique in momentum space as the one used in the present work may be therefore useful in this case as well. We hope to confirm this in a future work.

\clearpage
\appendix

\section{Matrix elements of the 1D hydrogen-like Dirac Hamiltonian}
\label{app:extension}

In this appendix, we argue that the matrix elements of the 1D hydrogen-like Dirac Hamiltonian $\bm{D}_Z$ can be defined on a set of functions larger than its domain.

We start from the definition of the Hamiltonian $\bm{D}_Z$ as having the same action of the free Dirac Hamiltonian $\bm{D}_0$ for $x\neq 0$, i.e.
\begin{eqnarray}
\bm{D}_Z \bm{\psi} = \bm{D}_0 \bm{\psi} \quad \text{on}\; \mathbb{R}\!\setminus\!\{0\}.
\label{}
\end{eqnarray}
with the $Z$-dependent domain
\begin{eqnarray}
\text{Dom}(\bm{D}_Z) = \left\{ \bm{\psi} \in H^1(\mathbb{R}\! \setminus \! \{0\},\mathbb{C}^2) \; |\; \bm{\psi}(0^+) = \b{M}_Z \bm{\psi}(0^-) \right\},
\label{}
\end{eqnarray}
where the matrix $\b{M}_Z$ is defined in Eq.~(\ref{1stboundarycond}). Let us consider two functions $\bm{\psi}$ and $\bm{\phi}$ in $\text{Dom}(\bm{D}_Z)$. Since they have a discontinuity at $x=0$, they can be written as, for $x \in \mathbb{R}$,
\begin{eqnarray}
\bm{\psi}(x) = \bm{\psi}_-(x) + H(x) \left[ \bm{\psi}_+(x) - \bm{\psi}_-(x) \right] \;\;\text{and}\;\;
\bm{\phi}(x) = \bm{\phi}_-(x) + H(x) \left[ \bm{\phi}_+(x) - \bm{\phi}_-(x) \right],
\end{eqnarray}
where $H$ is the Heaviside-step function, and $\bm{\psi}_\pm$ and $\bm{\phi}_\pm$ are (non-unique) functions in $H^1(\mathbb{R},\mathbb{C}^2)$ such that $\bm{\psi}_{-} =\bm{\psi}$ and $\bm{\phi}_{-} =\bm{\phi}$ on $\mathbb{R}^-$ and $\bm{\psi}_{+} =\bm{\psi}$ and $\bm{\phi}_{+} =\bm{\phi}$ on $\mathbb{R}^+$. Since they are chosen in $H^1(\mathbb{R},\mathbb{C}^2)$, the functions $\bm{\psi}_\pm$ and $\bm{\phi}_\pm$ are continuous at $x=0$ and we have $\bm{\psi}_\pm(0) = \bm{\psi}(0^\pm)$ and $\bm{\phi}_\pm(0) = \bm{\phi}(0^\pm)$. The matrix element $\braket{\bm{\phi}}{\bm{D}_Z \bm{\psi}}$ can be written as
\begin{eqnarray}
\braket{\bm{\phi}}{\bm{D}_Z \bm{\psi}} = \int_{-\infty}^{0} \bm{\phi}_-^\dagger(x) \bm{D}_0 \bm{\psi}_-(x) \d x + \int_{0}^{\infty} \bm{\phi}_+^\dagger(x) \bm{D}_0 \bm{\psi}_+(x) \d x.
\label{phiDpsibroken}
\end{eqnarray}
We would like to rewrite Eq.~(\ref{phiDpsibroken}) as a single integral over $(-\infty,+\infty)$ of $\bm{\phi}^\dagger(x) \bm{D}_0 \bm{\psi}(x)$ but the problem is that $\bm{\phi}^\dagger(x) \bm{D}_0 \bm{\psi}(x)$ formally contains a multiplication of distributions, namely the product of a Heaviside-step distribution and a Dirac-delta distribution, which is meaningless in the standard theory of distributions (where only the product of a distribution by a smooth function is defined). One way around this, in the spirit of Colombeau's generalized functions~\cite{Col-BOOK-92,GroKunObeSte-BOOK-01,Gsp-EJP-09,FilLorBan-JPA-12}, is to introduce regularized functions $\bm{\psi}_\epsilon$ and $\bm{\phi}_\epsilon$ depending on a regularization parameter $\epsilon>0$
\begin{eqnarray}
\bm{\psi}_\epsilon(x) = \bm{\psi}_-(x) + H_\epsilon(x) \left[ \bm{\psi}_+(x) - \bm{\psi}_-(x) \right]
\;\;\text{and}\;\;
\bm{\phi}_\epsilon(x) = \bm{\phi}_-(x) + H_\epsilon(x) \left[ \bm{\phi}_+(x) - \bm{\phi}_-(x) \right],
\label{psiepsphieps}
\end{eqnarray}
where $H_\epsilon(x)$ is a regularized Heasivide-step function obtained by convoluting $H$ with $\eta_\epsilon:  x \mapsto (1/\epsilon) \eta(x/\epsilon)$
\begin{eqnarray}
H_\epsilon(x) = (H * \eta_\epsilon)(x) 
= \int_{-\infty}^{+\infty} H(x-y) \frac{1}{\epsilon}\eta\left(\frac{y}{\epsilon}\right) \d y,
\end{eqnarray}
where $\eta$ is a mollifier, i.e. a compactly supported smooth real function satisfying $\int_{-\infty}^\infty \eta(x) \d x = 1$. The derivative of $\bm{\psi}_\epsilon$ is then
\begin{eqnarray}
\bm{\psi}_\epsilon'(x) &=& \bm{\psi}_-'(x) + H_\epsilon(x) \left[ \bm{\psi}_+'(x) - \bm{\psi}_-'(x) \right] 
+ \delta_\epsilon(x) \left[ \bm{\psi}_+(x) - \bm{\psi}_-(x) \right],
\end{eqnarray}
where $\delta_\epsilon = H_\epsilon'$ is the corresponding regularized Dirac-delta function
\begin{eqnarray}
\delta_\epsilon(x) = (\delta * \eta_\epsilon)(x) = \frac{1}{\epsilon}\eta\left(\frac{x}{\epsilon}\right).
\end{eqnarray}
Obviously, the limit $\epsilon \to 0$ corresponds to the non-regularized case with $\lim_{\epsilon\to 0} H_\epsilon = H$ and $\lim_{\epsilon\to 0} \delta_\epsilon = \delta$ in the sense of distributions. Now, since the regularized functions $\bm{\psi}_\epsilon$ and $\bm{\phi}_\epsilon$ are in $H^1(\mathbb{R},\mathbb{C}^2)$, we can integrate $\bm{\phi}^\dagger_\epsilon(x) \bm{D}_0 \bm{\psi}_\epsilon(x)$ over $(-\infty,+\infty)$ and take the limit $\epsilon \to 0$
\begin{eqnarray}
\lim_{\epsilon \to 0} \; \int_{-\infty}^\infty \bm{\phi}^\dagger_\epsilon(x) \bm{D}_0 \bm{\psi}_\epsilon(x) \d x = \braket{\bm{\phi}}{\bm{D}_Z \bm{\psi}} + S,
\label{}
\end{eqnarray}
where the regular part (not involving a delta function) of the integrand gives the value $\braket{\bm{\phi}}{\bm{D}_Z \bm{\psi}}$ in Eq.~(\ref{phiDpsibroken}) and the singular part (involving a delta function) of the integrand gives an additional contribution 
\begin{eqnarray}
S = \lim_{\epsilon \to 0} \; \int_{-\infty}^\infty -\i c \Biggl[ \bm{\phi}_-^\dagger(x) + H_\epsilon(x) \left[ \bm{\phi}_+^\dagger(x) - \bm{\phi}_-^\dagger(x) \right] \Biggl]
\bm{\sigma}_1 \delta_\epsilon(x)\left[ \bm{\psi}_+(x) - \bm{\psi}_-(x) \right] \d x .
\label{S}
\end{eqnarray}
The difficult term in this expression is the one involving the product $H_\epsilon(x)\delta_\epsilon(x)$ that we calculate now for any continuous and integrable function $f:\mathbb{R} \to \mathbb{C}$
\begin{eqnarray}
\lim_{\epsilon \to 0} \; \int_{-\infty}^\infty H_\epsilon(x)\delta_\epsilon(x) f(x)\d x  &=& \lim_{\epsilon \to 0} \; \int_{-\infty}^\infty \int_{-\infty}^\infty H(x-y) \frac{1}{\epsilon}\eta\left(\frac{y}{\epsilon}\right) \frac{1}{\epsilon}\eta\left(\frac{x}{\epsilon}\right) f(x)\d x \d y
\nonumber\\
&=&
\lim_{\epsilon \to 0} \; \int_{-\infty}^\infty \int_{-\infty}^\infty H(x-y)\eta(y) \eta(x) f(\epsilon x)\d x \d y
\nonumber\\
&=&
a f(0),
\label{Hepsdeltaeps}
\end{eqnarray}
where the real number $a$ is given by
\begin{eqnarray}
a = \int_{-\infty}^\infty \int_{-\infty}^\infty H(x-y) \eta(x) \eta(y) \d x \d y 
&=& \int_{-\infty}^\infty \int_{-\infty}^\infty [1 - H(y-x)] \eta(x) \eta(y) \d x \d y 
\nonumber\\
&=& 1 - \int_{-\infty}^\infty \int_{-\infty}^\infty H(y-x) \eta(x) \eta(y) \d x \d y
\nonumber\\
&=& 1 - a,
\label{}
\end{eqnarray}
i.e., $a=1/2$ independently of the mollifier $\eta$. Thus, we have shown that $\lim_{\epsilon\to 0} H_\epsilon \delta_\epsilon = (1/2) \delta$ in the sense of distributions, which corresponds to the result obtained in Colombeau theory. Notice that the value $a=1/2$ is obtained only if we use the same regularized Heasivide-step function $H_\epsilon$ for both $\bm{\psi}_\epsilon$ and $\bm{\phi}_\epsilon$ in Eq.~(\ref{psiepsphieps}). Using Eq.~(\ref{Hepsdeltaeps}), the term $S$ in Eq.~(\ref{S}) simplifies to
\begin{eqnarray}
S &=& -\frac{\i c}{2} \left[ \bm{\phi}_+^\dagger(0) + \bm{\phi}_-^\dagger(0) \right] 
\bm{\sigma}_1 \left[ \bm{\psi}_+(0) - \bm{\psi}_-(0) \right]
\nonumber\\
  &=& -\frac{\i c}{2} \left[ \bm{\phi}^\dagger(0^+) + \bm{\phi}^\dagger(0^-) \right] 
\bm{\sigma}_1 \left[ \bm{\psi}(0^+) - \bm{\psi}(0^-) \right]. \;\;
\end{eqnarray}
Using now the boundary condition $\bm{\psi}(0^+)= \b{M}_Z \bm{\psi}(0^-)$ and the matrix identity $-\i c \bm{\sigma}_1 \left[\b{M}_Z - \b{I}_2 \right]= (Z/2) \left[\b{M}_Z + \b{I}_2 \right]$, we can write $S$ as
\begin{eqnarray}
S  &=& -\frac{\i c}{2} \left[ \bm{\phi}^\dagger(0^+) + \bm{\phi}^\dagger(0^-) \right] 
\bm{\sigma}_1 \left[ \b{M}_Z  - \b{I}_2 \right] \bm{\psi}(0^-)
\nonumber\\
&=&  \frac{Z}{4} \left[ \bm{\phi}^\dagger(0^+) + \bm{\phi}^\dagger(0^-) \right] \left[ \b{M}_Z  + \b{I}_2 \right] \bm{\psi}(0^-)
\nonumber\\
&=&  \frac{Z}{4} \left[ \bm{\phi}^\dagger(0^+) + \bm{\phi}^\dagger(0^-) \right] \left[ \bm{\psi}(0^+) + \bm{\psi}(0^-) \right]
\nonumber\\
&=&  Z \bar{\bm{\phi}}^\dagger(0) \bar{\bm{\psi}}(0),
\end{eqnarray}
where we have defined $\bar{\bm{\psi}}(0) = [\bm{\psi}(0^+) + \bm{\psi}(0^-)]/2$ and $\bar{\bm{\phi}}(0) = [\bm{\phi}(0^+) + \bm{\phi}(0^-)]/2$. Therefore, we arrive at the following expression for $\braket{\bm{\phi}}{\bm{D}_Z \bm{\psi}}$
\begin{eqnarray}
\braket{\bm{\phi}}{\bm{D}_Z \bm{\psi}} = \lim_{\epsilon \to 0} \; \int_{-\infty}^\infty \bm{\phi}^\dagger_\epsilon(x) \bm{D}_0 \bm{\psi}_\epsilon(x) \d x
- Z \bar{\bm{\phi}}^\dagger(0) \bar{\bm{\psi}}(0).
\label{}
\end{eqnarray}
If we define the matrix element $\braket{\bm{\phi}}{\bm{D}_0 \bm{\psi}}$ using the regularized functions, or equivalently in Fourier space,
\begin{eqnarray}
\braket{\bm{\phi}}{\bm{D}_0 \bm{\psi}} = \lim_{\epsilon \to 0} \; \int_{-\infty}^\infty \bm{\phi}^\dagger_\epsilon(x) \bm{D}_0 \bm{\psi}_\epsilon(x) \d x 
= \int_{-\infty}^\infty \hat{\bm{\phi}}^\dagger(k) [  c \bm{\sigma}_1 \; k + \bm{\sigma}_3 \; mc^2 ] \hat{\bm{\psi}}(k) \d k,
\label{}
\end{eqnarray}
where $\hat{\bm{\psi}}$ and $\hat{\bm{\phi}}$ are the Fourier transforms of $\bm{\psi}$ and $\bm{\phi}$, respectively, we can finally write $\braket{\bm{\phi}}{\bm{D}_Z \bm{\psi}}$ as
\begin{eqnarray}
\braket{\bm{\phi}}{\bm{D}_Z \bm{\psi}} = \braket{\bm{\phi}}{\bm{D}_0 \bm{\psi}} 
- Z \bar{\bm{\phi}}^\dagger(0) \bar{\bm{\psi}}(0).
\label{phiDpsiapp}
\end{eqnarray}
Thus, the potential term in Eq.~(\ref{phiDpsiapp}) corresponds to interpreting the multiplication of the delta function $\delta (x)$ with a function $\bm{\psi}(x)$ discontinuous at $x=0$ as $\delta(x) \bm{\psi}(x) = \bar{\bm{\psi}}(0) \delta(x)$, as done in Refs.~\onlinecite{SubBha-JPC-72,Lap-AJP-83,FilLorBan-JPA-12}. This can also be understood from a distribution theory for discontinuous test functions~\cite{KurBom-PAMS-98}. Using the bound-state eigenfunction $\bm{\psi}_\text{b}^Z$ in Eq.~(\ref{psi0tilde}), it can be explicitly checked that calculating $\braket{\bm{\psi}_\text{b}^Z}{\bm{D}_Z \bm{\psi}_\text{b}^Z}$ using Eq.~(\ref{phiDpsiapp}) correctly gives the bound-state energy $\varepsilon_\text{b}^Z$ in Eq.~(\ref{epsilon0tilde}). We argue that Eq.~(\ref{phiDpsiapp}) can be used to define the matrix elements of the Hamiltonian $\bm{D}_Z$ on a $Z$-independent set of functions larger than its domain, and in particular containing continuous functions.

The formal Dirac Hamiltonian with a delta-function potential in Eq.~(\ref{Dv}) has also been interpreted in various works~\cite{CalKiaNog-AJP-87,CouNog-PRA-87,MckSte-PRC-87,AloDev-IJTP-00,GuiMunPirSan-FP-19} as another self-adjoint operator $\bm{D}_{\zeta}'$ corresponding to another boundary condition which has the same form as Eq.~(\ref{1stboundarycond}) but with $\theta$ replaced by $\theta'= \zeta/c$ for $0 \leq \zeta < c \pi$. In this case, Eq.~(\ref{phiDpsiapp}) becomes
\begin{eqnarray}
\braket{\bm{\phi}}{\bm{D}_\zeta' \bm{\psi}} = \braket{\bm{\phi}}{\bm{D}_0 \bm{\psi}} 
- 2c \tan\left(\frac{\zeta}{2c}\right) \bar{\bm{\phi}}^\dagger(0) \bar{\bm{\psi}}(0).
\label{phiDppsiapp}
\end{eqnarray}
This operator $\bm{D}_\zeta'$ is thus identical to the operator $\bm{D}_{Z(\zeta)}$ considered in the present work with the nuclear charge $Z(\zeta)= 2c \tan(\zeta/2c)$. The reason why the operator $\bm{D}_\zeta'$ is sometimes considered to be a realization of Eq.~(\ref{Dv}) is that, if we take a family of local regular real-valued potentials $v_\epsilon(x) = v(x/\epsilon)/\epsilon$ with $\int_{-\infty}^{\infty} v(x) \d x =1$ that converges (in the distribution sense) to the Dirac-delta function $\delta(x)$ as $\epsilon \to 0$, then the family of Dirac operators $\bm{D}_{\zeta,\epsilon}^\text{local}$ with these local potentials $-\zeta v_\epsilon$ converges (in the norm-resolvent sense) to $\bm{D}_\zeta'$ as $\epsilon \to 0$~\cite{MckSte-PRA-87,MckSte-PRC-87,CalKiaNog-AJP-87,Seb-LMP-89,Hug-LMP-95,Hug-RMP-97,Hug-JMAA-99,Tus-LMP-20}
\begin{eqnarray}
\bm{D}_{\zeta,\epsilon}^\text{local} = \bm{D}_{0} - \zeta v_\epsilon \underset{\epsilon \to 0}{\longrightarrow} \bm{D}_\zeta' = \bm{D}_{Z(\zeta)}.
\label{}
\end{eqnarray}
The fact that $\bm{D}_{\zeta,\epsilon}^\text{local}$ does not converge to $\bm{D}_{\zeta}$ but to $\bm{D}_{Z(\zeta)}$ is surprising. Instead, if we consider a family of Dirac operators $\bm{D}_{Z,\epsilon}^\text{nonlocal}$ with the nonlocal potentials $- Z \ket{v_\epsilon}\bra{v_\epsilon}$, where $v_\epsilon(x)$ still converges to the Dirac-delta function $\delta(x)$ as $\epsilon \to 0$, then $\bm{D}_{Z,\epsilon}^\text{nonlocal}$ converges (in the norm-resolvent sense), as expected, to $\bm{D}_Z$ as $\epsilon \to 0$~\cite{SutMat-PRA-81,CalKiaNog-AJP-87,CalKiaNog-PRC-88,Seb-LMP-89,Hug-LMP-95,KurBom-PAMS-98,HerTus-JMAA-22}
\begin{eqnarray}
\bm{D}_{Z,\epsilon}^\text{nonlocal} = \bm{D}_{0} - Z \ket{v_\epsilon}\bra{v_\epsilon} \underset{\epsilon \to 0}{\longrightarrow} \bm{D}_Z.
\label{DZepsnonloc}
\end{eqnarray}
In a finite basis $\{ \bm{\chi}_n \}$ of functions that are continuous at $x=0$, the matrix elements of the local and nonlocal operators $\bm{D}_{Z,\epsilon}^\text{local}$ and $\bm{D}_{Z,\epsilon}^\text{nonlocal}$ both reduce, in the limit $\epsilon \to 0$, to the matrix elements of the operator $\bm{D}_{Z}$ as given in Eq.~(\ref{phiDpsiapp}), i.e. 
\begin{eqnarray}
\lim_{\epsilon \to 0}\braket{\bm{\chi}_n}{\bm{D}_{Z,\epsilon}^\text{local} \bm{\chi}_m} = \lim_{\epsilon \to 0} \braket{\bm{\chi}_n}{\bm{D}_{Z,\epsilon}^\text{nonlocal} \bm{\chi}_m} = \braket{\bm{\chi}_n}{\bm{D}_{0} \bm{\chi}_m} - Z \bm{\chi}_n^\dagger(0) \bm{\chi}_m(0) = \braket{\bm{\chi}_n}{\bm{D}_{Z} \bm{\chi}_m}.
\label{}
\end{eqnarray}

\section{Green function from an asymmetric UV momentum cutoff}
\label{app:asymmcutoff}

In this appendix, we calculate the Green function that would have been obtained if we had used an asymmetric UV momentum cutoff in Section~\ref{sec:GeneralGreen}.

We consider a momentum interval $I(\Lambda)=[k_1(\Lambda),k_2(\Lambda)]$ containing $0$ and such that $\lim_{\Lambda \to \infty} k_1(\Lambda) = -\infty$ and $\lim_{\Lambda \to \infty} k_2(\Lambda) = +\infty$, but which is not necessarily symmetric around $0$, e.g. $I(\Lambda)=[-\Lambda,2\Lambda]$. The calculation in Section~\ref{sec:GeneralGreen} corresponds to using the symmetric interval $I(\Lambda)=[-\Lambda,\Lambda]$. Following the same steps as in the calculation in Section~\ref{sec:GeneralGreen}, we obtain for the variation of the Green function
\begin{eqnarray}
\Delta \bm{G}_Z(p,p';\omega) = - \frac{Z}{2\pi} \bar{\bm{G}}_0(p;\omega) \left[ \b{I}_2 + \frac{Z}{2\pi} \bar{\bar{\bm{G}}}_0(\omega) \right]^{-1}\bar{\bm{G}}_0(p';\omega),
\label{}
\end{eqnarray}
where $\bar{\bm{G}}_0(p;\omega)$ is given in Eq.~(\ref{barG0pomega}) and $\bar{\bar{\bm{G}}}_0(\omega) = \int_{I(\Lambda)} \bar{\bm{G}}_0(p;\omega) \d p$. In the limit $\Lambda \to \infty$, we obtain
\begin{eqnarray}
\lim_{\Lambda\to\infty}\bar{\bar{\bm{G}}}_0(\omega)
= \frac{\pi}{c} \left(\begin{array}{cc}
-g(\omega) & a  \\
a & g(-\omega)
\end{array}\right),
\label{}
\end{eqnarray}
with
\begin{eqnarray}
a  = \frac{c}{\pi} \lim_{\Lambda\to\infty} \int_{I(\Lambda)} \frac{c p}{\omega^2-\varepsilon_p^2} \d p.
\label{}
\end{eqnarray}
Depending on the choice of $I(\Lambda)$, the constant $a$ can be infinite or a finite real number independent of $\omega$. To see this, we note that the off-diagonal elements of $\bar{\bm{G}}_0(p;\omega)$ can be written as
\begin{eqnarray}
\frac{c p}{\omega^2-\varepsilon_p^2}  = - \frac{1}{c p} H(|p|-1) + R(p,\omega),
\label{}
\end{eqnarray}
where $H$ is the Heaviside-step function and $R$ is an odd function in $p$ and integrable in $p$ over $(-\infty,+\infty)$. Thus, in the limit $\Lambda \to \infty$, the integral of $R(p,\omega)$ over $p$ vanishes, and it remains
\begin{eqnarray}
a  &=& -\frac{1}{\pi} \lim_{\Lambda\to\infty} \int_{I(\Lambda)} \frac{1}{p} H(|p|-1)  \d p
\nonumber\\
&=& -\frac{1}{\pi} \lim_{\Lambda\to\infty} \ln \left( \frac{k_2(\Lambda)}{-k_1(\Lambda) } \right).
\label{}
\end{eqnarray}
For example, for the asymmetric interval $I=[-\Lambda,2\Lambda]$, we have $a=-(\ln 2)/\pi$. For a symmetric interval, i.e. $k_2(\Lambda)=-k_1(\Lambda)$, we recover $a=0$. 

Thus, we obtain, in the limit $\Lambda \to \infty$, a Green function with an arbitrary parameter $a \in \mathbb{R}$, whose value depends on how the infinite UV momentum limit is taken. In position space, we can extract from this Green function its associated boundary-condition matrix at $x = 0$
\begin{eqnarray}
\b{M}_Z^a = 
\frac{1}{1-(a-\i)^2 \lambda^2} \left(\begin{array}{cc}
1 - (a^2+1) \lambda^2 & 2\,\i \lambda  \\
2\,\i \lambda & 1 - (a^2+1) \lambda^2
\end{array}\right),
\label{}
\end{eqnarray}
with $\lambda = Z/2c$. This matrix can be rewritten in the form (see Ref.~\onlinecite{BenDab-LMP-94}) 
\begin{eqnarray}
\b{M}_Z^a = 
w \left(\begin{array}{cc}
A & \i\, B  \\
-\i\, C & D
\end{array}\right),
\label{}
\end{eqnarray}
where $w=(1-(a+\i)^2\lambda^2)/(1-(a-\i)^2\lambda^2)$, $A=D=(1 - (a^2+1) \lambda^2)/\sqrt{P}$, $B=-C=2 \lambda/\sqrt{P_a}$, and $P=1-2(a^2-1)\lambda^2 + (a^2+1)^2 \lambda^4$. For $a=0$, we recover the boundary condition in Eq.~(\ref{DomD}). For a general parameter $a$, it can be checked that $|w|=1$, $AD-BC=1$, and $A,B,C,D \in \mathbb{R}$, so that it still corresponds to a self-adjoint Dirac Hamiltonian $D_Z^a$ with a point-interaction potential~\cite{BenDab-LMP-94}. More precisely, this Hamiltonian $D_Z^a$ is (see Refs.~\onlinecite{CouNogFer-JPA-97,CouNogTom-JPA-99a,AlbFeiKur-RMP-04} for related discussions)
\begin{eqnarray}
\bm{D}_Z^a = \b{U} \bm{D}_{Z_a} \b{U}^\dagger,
\label{}
\end{eqnarray}
with the unitary operator $\b{U} = \chi \; \b{I}_2$ where $\chi$ is the function of modulus $1$ given by 
\begin{eqnarray}
\chi(x)  = 
\begin{cases}
  1, &  x<0\\
  w, & x \geq 0,
\end{cases}
\label{}
\end{eqnarray}
and $Z_a/2c = \sqrt{(1-A)/(1+A)}$. This Hamiltonian $D_Z^a$ is thus unitarily equivalent to the Hamiltonian in Eqs.~(\ref{Dpsi=D0psi})-(\ref{DomD}) but with the nuclear charge $Z$ replaced by $Z_a$. This shows the importance of choosing a symmetric UV momentum cutoff to obtain the Green function corresponding to the Hamiltonian in Section~\ref{sec:hamiltonian} with the correct nuclear charge $Z$.

\section{Error in the calculation of the vacuum-polarization density in position space}
\label{app:error}
In this appendix, we explain why the Dirac-delta contribution was missed in the calculation of the vacuum-polarization density in position space in Appendix D of Ref.~\onlinecite{AudTou-JCP-23}. The error was to conclude from the expression of the vacuum-polarization density matrix in position space in Eq.~(D10) of Ref.~\onlinecite{AudTou-JCP-23}, 
\begin{align}
  \b{n}^{\text{vp}}_{1}(x,x') = \frac{1}{2\pi}\int_{-\infty}^{\infty}  \Delta\b{G}(x,x';\i u)\d u
\end{align}
where $\Delta\b{G}(x,x';iu)$ is the  variation of the position-space Green function,
that the position-space vacuum-polarization density was given by the expression in Eq.~(D11) of Ref.~\onlinecite{AudTou-JCP-23},
\begin{align}
  ``{n}^{\text{vp}}(x) =  \frac{1}{2\pi} \int_{-\infty}^{\infty} \tr[\Delta\b{G}(x,x;\i u)]\d u \text{''}.
\end{align}
Indeed, while $\Delta\b{G}(x,x';\i u)$ is continuous around the diagonal $x=x'$ for a given $u$, the integral over $u$ contains a singular contribution which is
not continuous at $x=x'=0$. This singular term in $\tr[ \b{n}^{\text{vp}}_{1}(x,x')]$, at first order in $Z$ for simplicity, is
\begin{eqnarray}
  \tr[\b{n}^{\text{vp},(1)}_{1,\text{sing}}(x,x')] &=& - \frac{Z}{4\pi c^2} [\sgn(x)\sgn(x') - 1] \int_{-\infty}^\infty e^{-\kappa(\i u) (|x|+|x'|)} \d u
  \nonumber\\
  &=&  - \frac{m Z}{2\pi} [\sgn(x)\sgn(x') - 1]  \; K_1\Big(mc \;(|x|+|x'|)\Big),
\end{eqnarray}
where $K_1$ is the modified Bessel function of the second kind. This term has a diagonal $x=x'$ that vanishes for $x\neq 0$, but gives raise of the Dirac-delta contribution in the Uehling vacuum-polarization density $n^{\text{vp},(1)}(x)$. To see this, we first use $K_1(z)\sim 1/z$ as $z\to0$ to find
\begin{eqnarray}
 \tr[\b{n}^{\text{vp},(1)}_{1,\text{sing}}(x,x')] &\isEquivTo{x,x'\to 0}&   \frac{Z}{2\pi c} \frac{[1-\sgn(x)\sgn(x')]}{|x|+|x'|}.
\end{eqnarray}
In the spirit of Refs.~\onlinecite{Bri-PAMS-88,Bri-PJM-91}, we define the $x=x'$ diagonal of the kernel $\tr[\b{n}^{\text{vp},(1)}_{1,\text{sing}}(x,x')]$ by averaging it around the diagonal as
\begin{eqnarray}
  n^{\text{vp},(1)}_\text{sing}(x) =  \lim_{\epsilon \to 0^+} \frac{1}{(2\epsilon)^2} \int_{x-\epsilon}^{x+\epsilon}  \int_{x-\epsilon}^{x+\epsilon} \tr[\b{n}^{\text{vp},(1)}_{1,\text{sing}}(y,y')] \d y \d y'.
\label{nvp1singx}
\end{eqnarray}
This gives
\begin{eqnarray}
  n^{\text{vp},(1)}_\text{sing}(x) = \frac{Z}{\pi c} \lim_{\epsilon \to 0^+}  \frac{f(x/\epsilon)}{\epsilon},
  \label{nvp1singlim}
\end{eqnarray}
where
\begin{eqnarray}
  f(x) = 
  \begin{cases}
  (1/2) \ln(4/(1-x^2)) - x \arctanh(x), &  |x|<1\\
  0, & |x|\geq 1,
  \end{cases}
  \end{eqnarray}
and $\int_{-\infty}^{\infty} f(x) \d x = 1$. Thus, in the sense of distributions, the limit in Eq.~(\ref{nvp1singlim}) tends to a Dirac-delta function
\begin{eqnarray}
  n^{\text{vp},(1)}_\text{sing}(x) = \frac{Z}{\pi c} \delta(x),
\end{eqnarray}
in agreement with Eq.~(\ref{nvp1x}). The derivation in momentum space presented in Section~\ref{sec:vpd} of this paper is valid without the need of a regularization like in Eq.~(\ref{nvp1singx}), because $\tr[\hat{\b{n}}^{\text{vp},(1)}_{1}(p,p')]$ is smooth in $(p,p')$.

\section{Rate of convergence of the bound-state energy in a plane-wave basis}
\label{app:rate}

In this appendix, we study the rate of convergence of the bound-state energy of the 1D hydrogen-like Dirac atom in a plane-wave basis.

The exact bound-state eigenfunction of the 1D hydrogen-like Dirac Hamiltonian $\b{D}_Z$ has the form (see Eq.~(\ref{psi0tilde}))
\begin{eqnarray}
\bm{\psi}_\text{b}^Z(x) = \left(\begin{array}{c}
\psi^\text{L}(x)\\
\psi^\text{S}(x)
\end{array}\right), 
\end{eqnarray}
with $\psi^\text{L}(x) = A_\text{b} e^{-\kappa_\text{b} |x|}$ and  $\psi^\text{S}(x) = A_\text{b} \i \sgn(x) (Z/2c) e^{-\kappa_\text{b} |x|}$ (the expressions of the constants $A_\text{b}$ and $\kappa_\text{b}$ are given after Eq.~(\ref{psi0tilde})). The associated exact bound-state energy is
\begin{eqnarray}
\varepsilon_\text{b}^Z &=&  \braket{\bm{\psi}_\text{b}^Z}{\bm{D}_Z \bm{\psi}_\text{b}^Z} 
\nonumber\\
&=&  \braket{\bm{\psi}_\text{b}^Z}{mc^2\bm{\sigma}_3 \bm{\psi}_\text{b}^Z}.
\label{virial}
\end{eqnarray}
The last equality in Eq.~(\ref{virial}) comes from the relativistic virial theorem~\cite{Bra-PRD-83,Sha-INC-03,FroGod-A-22}, which can be shown, e.g. using a scaling argument, to have the same form for the 1D Dirac equation with a Dirac-delta potential as for the 3D Dirac equation with a Coulomb potential. It provides a convenient expression for calculating the energy.

We consider a complete plane-wave basis $\{ \chi_n \}_{n\in \mathbb{Z}}$ on the interval $[-L/2,L/2]$ where $\chi_n(x) = (1/\sqrt{L}) e^{\i k_n x}$ and $k_n = 2\pi n/L$, and use it to expand the restriction of the exact bound-state eigenfunction $\bm{\psi}_\text{b}^Z$ to the interval $[-L/2,L/2]$. Since $\psi^\text{L}$ and $\psi^\text{S}$ are even and odd functions, respectively, we introduce even and odd basis functions
\begin{eqnarray}
\chi_n^\text{g}(x) = \begin{cases}
\chi_0(x) & \text{for} \; n=0\\[0.2cm]
\dfrac{\chi_n(x) + \chi_{-n}(x)}{\sqrt{2}} & \text{for} \; n\in \mathbb{N}^*
\end{cases} \;\;\text{and}\;\; \chi_n^\text{u}(x) = \dfrac{\chi_n(x) - \chi_{-n}(x)}{\sqrt{2}} & \text{for} \; n\in \mathbb{N}^*.
\end{eqnarray}
For $x \in [-L/2,L/2]$, we have thus
\begin{eqnarray}
\psi^\text{L}(x)=\sum_{n=0}^{\infty}c_n^\text{L} \; \chi_n^\text{g}(x) \;\;\text{and}\;\; \psi^\text{S}(x)=\sum_{n=1}^{\infty}c_n^\text{S}\;\chi_n^\text{u}(x),
\end{eqnarray}
with coefficients 
\begin{eqnarray}
c_n^\text{L}= \braket{\psi^\text{L}}{\chi_n^\text{g}}_{\!_L}  = \begin{cases}
\dfrac{2 A_\text{b}}{\kappa_\text{b}\sqrt{L}} \left( 1 -  e^{-\kappa_\text{b} L/2}  \right) & \text{for} \; n=0\\[0.3cm]
\dfrac{2\sqrt{2} A_\text{b} \kappa_\text{b}}{\sqrt{L}} \dfrac{1 - (-1)^n e^{-\kappa_\text{b} L/2}}{\kappa_b + k_n^2} & \text{for} \; n\in \mathbb{N}^*
\end{cases} \;\;\text{and}\;\; c_n^\text{S}= \braket{\psi^\text{S}}{\chi_n^\text{u}}_{\!_L}  = -\frac{Z k_n}{2c \kappa_\text{b}} c_n^\text{L} & \text{for} \; n\in \mathbb{N}^*.
\label{coeffs}
\end{eqnarray}
As $n\to\infty$, the large-component coefficients $c_n^\text{L}$ decay as $1/n^2$ while the small-component coefficients $c_n^\text{S}$ decay as $1/n$.

We now consider the best approximation (in the sense of the $L^2$ norm) $\tilde{\bm{\psi}}_\text{b}^Z$ to $\bm{\psi}_\text{b}^Z$ in the finite basis $\{ \chi_n \}_{|n| \leq n_\text{max}}$, for $x \in [-L/2,L/2]$,
\begin{eqnarray}
\tilde{\bm{\psi}}_\text{b}^Z(x) = \left(\begin{array}{c}
\tilde{\psi}^\text{L}(x)\\
\tilde{\psi}^\text{S}(x)
\end{array}\right), 
\end{eqnarray}
where
\begin{eqnarray}
\tilde{\psi}^\text{L}(x)=\sum_{n=0}^{n_\text{max}}c_n^\text{L}\; \chi_n^\text{g}(x)  \;\;\text{and}\;\; \tilde{\psi}^\text{S}(x)=\sum_{n=1}^{n_\text{max}}c_n^\text{S} \;\chi_n^\text{u}(x).
\end{eqnarray}
The corresponding approximation to the bound-state energy, as a function of the IR cutoff $L$ and the UV cutoff $\Lambda=2\pi n_\text{max}/L$, is
\begin{eqnarray}
\tilde{\varepsilon}_\text{b}^Z(L,\Lambda) &=& \braket{\tilde{\bm{\psi}}_\text{b}^Z}{mc^2\bm{\sigma}_3 \tilde{\bm{\psi}}_\text{b}^Z}_{\!_L}
 = mc^2 \left( \sum_{n=0}^{n_\text{max}} |c_n^\text{L} |^2 - \sum_{n=1}^{n_\text{max}} |c_n^\text{S}|^2 \right).
\label{}
\end{eqnarray}
From the expression of the coefficients in Eq.~(\ref{coeffs}), we find the behavior of the bound-state energy as $\Lambda \to \infty$
\begin{eqnarray}
\tilde{\varepsilon}_\text{b}^Z(L,\Lambda) \isEquivTo{\Lambda \to  \infty} \tilde{\varepsilon}_{\text{b}}^Z(L,\infty)  + \frac{A_\text{b}^2 Z^2 \left( 1 + e^{-\kappa_\text{b} L} \right) }{\pi c^2 \; \Lambda},
\label{}
\end{eqnarray}
with
\begin{eqnarray}
\tilde{\varepsilon}_{\text{b}}^Z(L,\infty) = \lim_{\Lambda \to  \infty} \tilde{\varepsilon}_\text{b}^Z (L,\Lambda) = \left( 1 - e^{-\kappa_\text{b} L} \right) \varepsilon_{\text{b}}^Z,
\label{}
\end{eqnarray}
where $\varepsilon_{\text{b}}^Z$ is the exact bound-state energy.
We thus find that $\tilde{\varepsilon}_\text{b}^Z(L,\Lambda)$ converges as $1/\Lambda$ as $\Lambda \to \infty$. This asymptotic convergence rate comes entirely from the small-component contribution which, having a discontinuity at $x=0$ in addition to a derivative discontinuity, represents the limiting factor in the convergence of the bound-state energy. We also see that, at least for large enough $\Lambda$, the bound-state energy converges exponentially with $L$ as $L\to\infty$. Even though the best approximate eigenfunction $\tilde{\bm{\psi}}_\text{b}^Z$ in the sense of the $L^2$ norm considered here does not exactly correspond to the approximate eigenfunction obtained in Section~\ref{sec:calculations} by diagonalizing the Hamiltonian in the plane-wave basis, in practice we expect a similar convergence rate for the latter case.


\section*{Acknowledgements}
We thank E. Séré for discussions. U.M. has received funding from the European Union's Horizon 2020 research and innovation programme under the Marie Sklodowska-Curie grant agreement N°945332.


\begin{thebibliography}{100}
\bibitem{AudMorLevTou-JJJ-XX-note1}
An equivalent alternative expression of $\b{n}^{\text{vp}}_\text{reg}(x)$ was
  also given in Eq. (66) of Ref.~\cite{AudTou-JCP-23}. In the latter equation,
  a $\sgn(x)$ function was missing in the off-diagonal elements of
  $\b{n}^{\text{vp}}_\text{reg}(x)$.
\end{thebibliography}

\end{document}